\documentclass[journal]{IEEEtran}

\usepackage{amsmath}
\usepackage{amssymb}
\usepackage{amsmath}
\usepackage{amssymb}
\usepackage{epsfig}
\usepackage{epsf}
\usepackage{subfigure}
\usepackage{graphicx}
\usepackage{cite}
\usepackage{url}
\usepackage[]{authblk}

\def\BibTeX{{\rm B\kern-.05em{\sc i\kern-.025em b}\kern-.08em
    T\kern-.1667em\lower.7ex\hbox{E}\kern-.125emX}}

\newtheorem{theorem}{Theorem}

\newtheorem{remark}{Remark}

\begin{document}


\title{Two-User Erasure Interference Channels\\ with Local Delayed CSIT}


\author{Alireza~Vahid,
        Robert~Calderbank				
        \thanks{Alireza Vahid is with the information initiative at Duke (\emph{i}iD), Duke University, Durham, NC, USA. Email: {\sffamily alireza.vahid@duke.edu}.}
				\thanks{Robert Calderbank is with departments of Electrical and Computer Engineering, Mathematics, and Computer Science, Duke University, Durham, NC, USA. Email: {\sffamily robert.calderbank@duke.edu}.}
				\thanks{The work of A. Vahid and R. Calderbank was supported in part by AFOSR under award No FA 9550-13-1-0076.}
				\thanks{Preliminary parts of this work were presented at the 2015 International Symposium on Information Theory (ISIT)~\cite{FBBudgetISIT}.}
}

\maketitle


\begin{abstract}
We study the capacity region of two-user erasure interference channels with \emph{local} delayed channel state information at the transmitters. In our model, transmitters have local mismatched outdated knowledge of the channel gains. We propose a transmission strategy that only relies on the delayed knowledge of the {\it outgoing} links at each transmitter and achieves the outer-bound for the scenario in which transmitters learn the \emph{entire} channel state with delay. Our result reveals the subset of the channel state information that affects the capacity region the most. 

We also identify cases in which local delayed knowledge of the channel state does not provide any gain over the zero knowledge assumption. To do so, we revisit a long-known intuition about interference channels that as long as the marginal distributions at the receivers are conserved, the capacity remains the same. We take this intuition and impose a certain spatial correlation among channel gains such that the marginal distributions remain unchanged. Then we provide an outer-bound on the capacity region of the channel with correlation that matches the capacity region when transmitters do not have access to channel state information.
\end{abstract}



\section{Introduction}
\label{Section:Introduction}

\begin{figure}[ht]
\centering
\includegraphics[height = 4cm]{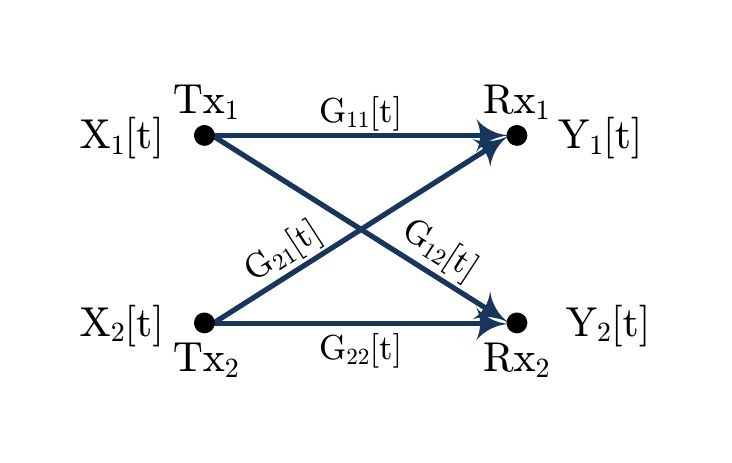}
\caption{Two-user Erasure Interference Channel.\label{Fig:detIC}}
\end{figure}

The canonical two-user interference channel (IC) introduced in~\cite{ahlswede1974capacity} is a fundamental building block in wireless communications and information theory. The behavior and the capacity of multi-terminal wireless networks could not be understood without a good grasp of the two-user interference channel. Subsequently there developed a significant body of work aimed at understanding the capacity region of this problem (\emph{e.g.},~\cite{sato1977two,HanKoba:it81,etkin2008gaussian}). Again, when it comes to fading interference channels, understanding the capacity of the canonical two-user IC is of great importance. Recent results~\cite{BFICAllerton,BFICISIT2012,AlirezaBFICDelayed,jeon2013capacity} address the capacity region of the canonical two-user fading IC under a specific channel distribution: the two-user erasure Interference Channel depicted in Fig.~\ref{Fig:detIC} where the channel gains at each time are drawn from the binary field according to some Bernoulli distributions. The input-output relation of this channel at time $t$ is given by
\begin{equation} 
Y_i[t] = G_{ii}[t] X_i[t] \oplus G_{\bar{i}i}[t] X_{\bar{i}}[t], \quad i = 1, 2,
\end{equation}
where $\bar{i} = 3 - i$, $G_{ii}[t], G_{\bar{i}i}[t] \in \{ 0, 1\}$, $X_i[t] \in \{ 0, 1\}$ is the transmit signal of transmitter $i$ at time $t$, and $Y_i[t] \in \{ 0, 1\}$ is the observation of receiver $i$ at time $t$. All algebraic operations are in $\mathbb{F}_2$. This model is a very good abstraction of wireless packet networks as discussed in~\cite{vahid2014communication}. 

In~\cite{AlirezaBFICDelayed}, the capacity region of the two-user erasure IC was characterized under the assumption of \emph{global} delayed channel state  information at the transmitters (CSIT) where each transmitter at time $t$ knows all the channel realizations up to time $\left( t - 1 \right)$. The result with global delayed CSIT includes novel transmission strategies and provides a new technique for deriving outer-bounds. However from the achievability perspective, the results rely strongly on the fact that transmitters learn the \emph{entire} channel state information (CSI) with delay. While this assumption might be justified  for the small canonical two-user IC, for large-scale networks such assumption might not be feasible at all. Thus, we aim to understand whether it is possible to achieve the same performance of global delayed CSIT with strictly smaller \emph{local} delayed CSIT. 

We consider several possible choices for the available delayed CSIT at each transmitter as shown in Fig.~\ref{Fig:Hierarchy} (such locality of knowledge was considered in~\cite{DavidKaoLocal} for a different setup). We demonstrate that it is sufficient for each transmitter to only have the knowledge of its outgoing channel gains with delay in order to achieve the same performance of global delayed CSIT. From this result we learn that each transmitter has to resolve the interference it creates at the unintended receiver. In other words, ``everyone should clean up their own mess!'' We identify those cases when local delayed CSIT provides no gain in capacity over the baseline where transmitters have no knowledge of CSI. Basically, we identify the most important subset or ``the most significant bits'' (MSBs) of the delayed channel state information.

Our contributions are thus multi-fold. We propose a new transmission strategy that solely relies on local delayed knowledge of the outgoing links at each transmitter. We show that this transmission strategy achieves the capacity region of the problem under the global delayed assumption. While this transmission strategy achieves the same region as the one proposed in~\cite{AlirezaBFICDelayed}, it has a much simpler structure and incorporates new ingredients. To be precise, the strategy of~\cite{AlirezaBFICDelayed} cannot be applied when transmitters have access to local delayed CSI. Our proposed scheme smartly takes advantage of the statistics of the channel to compensate for the lack of global knowledge. Moreover, the proposed scheme reduces the complexity of the transmission protocol by reducing the number of virtual queues it generates at each transmitter when compared to that of~\cite{AlirezaBFICDelayed}. Our result provides a better intuition and a deeper understanding of the coding opportunities that arise from delayed CSIT. 

In order to identify the cases where local delayed CSIT does not provide any gain over no CSIT assumption, we borrow the intuition provided by Sato~\cite{sato1977two}: ``the capacity region of all interference channels that have the same marginal distributions is the same.'' We take this intuition and create a certain spatial correlation among channel gains such that the marginal distributions remain unchanged. Then, we provide an outer-bound on the capacity region of the channel with correlation which can be achieved with no CSIT.

There is some prior work assessing the value of delayed CSIT. It was used in~\cite{jolfaei1993new} to create transmitted signals that are simultaneously useful for multiple users in a broadcast channel. These ideas were then extended to different wireless networks with delayed CSIT. Some examples are the study of erasure broadcast channels~\cite{georgiadis2009broadcast}, the DoF region of broadcast channels~\cite{maddah2012completely}, the approximate capacity region of Gaussian broadcast channels~\cite{vahid2013approximate,MISOBC2016TCOMM}, and the DoF region of multi-antenna multi-user Gaussian ICs and X channels~\cite{GhasemiX1,Vaze_DCSIT_MIMO_BC,Jafar_Retrospective,ghasemi2011interference}.  

The rest of the paper is organized as follows. In Section~\ref{Section:Problem}, we formulate our problem. In Section~\ref{Section:Main}, we present our main results. Sections~\ref{Section:NoGain} and~\ref{Section:FullGain} are dedicated to the proof of the main results. Next, in Section~\ref{Section:Discussion}, we discuss the implications of our results for more general settings. Section~\ref{Section:Conclusion} concludes the paper.


\section{Problem Setting}
\label{Section:Problem}

To study the capacity region of the two-user fading interference channels with local delayed CSIT, we consider an erasure model. In the erasure model, the channel gain from transmitter ${\sf Tx}_i$ to receiver ${\sf Rx}_j$ at time $t$ is the binary field element denoted by $G_{ij}[t] \in \{0,1\}$, $i,j \in \{1,2\}$. Channel gains are distributed as independent Bernoulli random variables (independent across time and space). We refer to this model as binary fading. We define the channel state information at time instant $t$ to be the set 
\begin{align}
G[t] \overset{\triangle}= \left \{ G_{11}[t], G_{12}[t], G_{21}[t], G_{22}[t] \right \}.
\end{align}
In this work, we focus on a homogeneous setting for simplicity of notations where
\begin{align}
G_{ij}[t] \overset{d}\sim \mathcal{B}(p), \qquad i,j = 1,2,
\end{align}
for $0 \leq p \leq 1$. Define $q \overset{\triangle}= 1 - p$.

In the sequel, we assume that each receiver has instantaneous knowledge of the channel state information. However, transmitter ${\sf Tx}_i$ will be aware of a subset of the CSI $\mathcal{S}_{{\sf Tx}_i}$ with unit delay, $i=1,2$. More precisely,
\begin{align}
\mathcal{S}_{{\sf Tx}_i} \subseteq \left \{ \left( 1, 1 \right), \left( 1, 2 \right), \left( 2, 1 \right), \left( 2, 2 \right)  \right \},
\end{align}
meaning that if $\left( k, \ell \right) \in \mathcal{S}_{{\sf Tx}_i}$, then ${\sf Tx}_i$ at time $t$ has access to $G_{k\ell}[1],G_{k\ell}[2],\ldots,G_{k\ell}[t-1]$ (or simply $G_{k\ell}^{t-1}$). Moreover, we consider the symmetric setting where we have
\begin{align}
\label{eq:symmetricS}
\left( k, \ell \right) \in \mathcal{S}_{{\sf Tx}_1} \Leftrightarrow \left( \bar{k}, \bar{\ell} \right) \in \mathcal{S}_{{\sf Tx}_2},  \qquad k,\ell = 1,2,
\end{align}
for $\bar{k} \overset{\triangle}= 3 - k$ and $\bar{\ell} \overset{\triangle}= 3 - \ell$.

\begin{figure*}[t]
\centering
\includegraphics[height = 12cm]{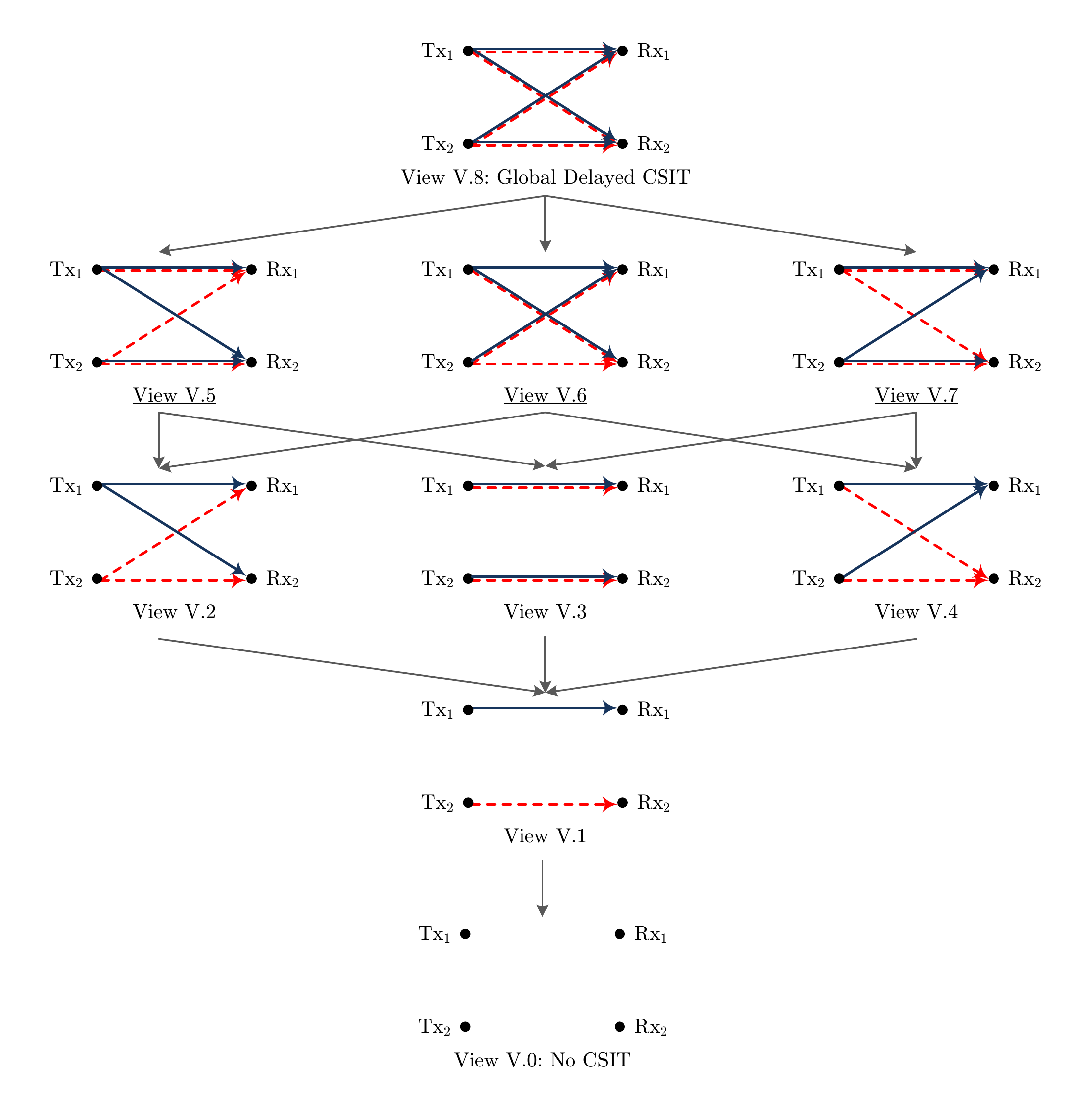}
\caption{\it Different choices of local delayed CSIT. Dark blue solid arrows denote the links for which ${\sf Tx}_1$ learns the channel gains with unit delay; while red dashed arrows denote the links for which ${\sf Tx}_2$ learns the channel gains with unit delay.\label{Fig:Hierarchy}}
\end{figure*}

We consider a total of $8$ possible choices of local delayed CSIT for each transmitter as follows.
\begin{itemize}

\item View V.0: This case would be our base-line that we refer to as no CSIT where
\begin{align}
\mathcal{S}_{{\sf Tx}_i} = \emptyset, \qquad i = 1, 2.
\end{align}

\item View V.1: In this case each transmitter is aware of channel value to its corresponding receiver with unit delay, \emph{i.e.}
\begin{align}
\mathcal{S}_{{\sf Tx}_i} = \{ \left( i, i \right) \}, \qquad i = 1, 2.
\end{align}

\item View V.2: In this case each transmitter is aware of channel value of the outgoing links with unit delay, \emph{i.e.}
\begin{align}
\mathcal{S}_{{\sf Tx}_i} = \{ \left( i, i \right), \left( i, \bar{i} \right) \}, \qquad i = 1, 2.
\end{align}

\item View V.3: In this case each transmitter is aware of channel value of the direct links with unit delay, \emph{i.e.}
\begin{align}
\mathcal{S}_{{\sf Tx}_i} = \{ \left( i, i \right), \left( \bar{i}, \bar{i} \right) \}, \qquad i = 1, 2.
\end{align}

\item View V.4: In this case each transmitter is aware of channel value of the links connected to its receiver with unit delay, \emph{i.e.}
\begin{align}
\mathcal{S}_{{\sf Tx}_i} = \{ \left( i, i \right), \left( \bar{i}, i \right) \}, \qquad i = 1, 2.
\end{align}

\item View V.5: In this case we have
\begin{align}
\mathcal{S}_{{\sf Tx}_i} = \{ \left( i, i \right), \left( i, \bar{i} \right), \left( \bar{i}, \bar{i} \right) \}, \qquad i = 1, 2.
\end{align}

\item View V.6: In this case we have
\begin{align}
\mathcal{S}_{{\sf Tx}_i} = \{ \left( i, i \right), \left( i, \bar{i} \right), \left( \bar{i}, i \right) \}, \qquad i = 1, 2.
\end{align}

\item View V.7: In this case we have
\begin{align}
\mathcal{S}_{{\sf Tx}_i} = \{ \left( i, i \right), \left( \bar{i}, i \right), \left( \bar{i}, \bar{i} \right) \}, \qquad i = 1, 2.
\end{align}

\item View V.8: Finally View 8 corresponds to the scenario where each transmitter is aware of the entire CSI with delay. We shall refer to this view as the global delayed CSIT where
\begin{align}
\mathcal{S}_{{\sf Tx}_i} = \{ \left( 1, 1 \right), \left( 1, 2 \right), \left( 2, 1 \right), \left( 2, 2 \right) \}, \qquad i = 1, 2.
\end{align}

\end{itemize}

Fig.~\ref{Fig:Hierarchy} pictorially depicts these $8$ different views. We note that the transmitters' knowledge in View V.0 is a subset of V.1; in V.1 is a subset of V.2, V.3, and V.4; in V.2 and V.3 is a subset of V.5; and so on. This hierarchical structure is also shown in Fig.~\ref{Fig:Hierarchy} using downward arrows\footnote{A similar set of local views for the channel state information was studied in~\cite{DavidKaoLocal} in the context of two-user Gaussian interference channel (not fading) to identify the views in which one could outperform TDMA.}.

We consider the scenario in which ${\sf Tx}_i$ wishes to reliably communicate message $\hbox{W}_i \in \{ 1,2,\ldots,2^{n R_i}\}$ to ${\sf Rx}_i$ during $n$ channel uses, $i = 1,2$. We assume that the messages and the channel gains are {\it mutually} independent and the messages are chosen uniformly. Let message $\hbox{W}_i$ be encoded as $X_i^n$ at transmitter ${\sf Tx}_i$ using the encoding function $f_i(\hbox{W}_i, \mathcal{S}_{{\sf Tx}_i})$, which depends on the available channel state information at the transmitter. Receiver ${\sf Rx}_i$ is only interested in decoding $\hbox{W}_i$, and it will decode the message using the decoding function $\widehat{\hbox{W}}_i = g_i(Y_i^n,G^n)$. An error occurs when $\widehat{\hbox{W}}_i \neq \hbox{W}_i$. The average probability of decoding error is given by
\begin{equation}
\label{eq:errorterms}
\lambda_{i,n} = \mathbb{E}[P[\widehat{\hbox{W}}_i \neq \hbox{W}_i]], \hspace{5mm} i = 1, 2,
\end{equation}
and the expectation is taken with respect to the random choice of the transmitted messages $\hbox{W}_1$ and $\hbox{W}_2$. 

A rate-tuple $(R_1,R_2)$ is said to be achievable, if there exists encoding and decoding functions at the transmitters and the receivers respectively, such that the decoding error probabilities $\lambda_{1,n},\lambda_{2,n}$ go to zero as $n$ goes to infinity for the given choice of $\mathcal{S}_{{\sf Tx}_1}$ and $\mathcal{S}_{{\sf Tx}_2}$. The capacity region for View V.$j$, \emph{i.e} 
\begin{align}
\mathcal{C}\left( \text{V.}j \right), \qquad j=1,2,\ldots,8,
\end{align}
is the closure of all achievable rate-tuples. In the following section, we present our main results.


\section{Statement of the Main Results}
\label{Section:Main}

Our main objective is to understand the ramification of {\emph local} delayed CSIT on the capacity region of two-user erasure Interference Channels. We establish the capacity region of two-user erasure ICs with no CSIT and global delayed CSIT as our benchmarks. Then we are interested in finding the answer to the following questions:

\begin{enumerate}

\item What is the minimum amount of delayed CSIT required to outperform no CSIT?

\item Is it possible to achieve the performance of global delayed CSIT with a strictly smaller subset of knowledge at each transmitter?

\end{enumerate}

Our work essentially identifies the ``MSBs'' of the delayed CSIT in two-user erasure ICs.

\subsection{Benchmarks}

Our base-line is the no CSIT scenario. In other words, the only available knowledge at the transmitters is the distribution from which the channel gains are drawn. In this case, it is easy to see that for any input distribution, the two received signals are \emph{statistically} the same. Therefore, the capacity region in this case is the same as the intersection of the capacity regions of the multiple-access channels (MACs) formed at the two receivers.  Thus, $\mathcal{C}\left( \text{V.}0 \right)$, is the set of all rate-tuples $\left( R_1, R_2 \right)$ satisfying
\begin{equation}
\label{eq:RegionNoCSIT}
\left\{ \begin{array}{ll}
\vspace{1mm} 0 \leq R_i \leq p, & i = 1,2, \\
R_1 + R_2 \leq 1-q^2. &
\end{array} \right.
\end{equation}

The other extreme point is the global delayed CSIT model. From~\cite{AlirezaBFICDelayed}, we know that the capacity region of the two-user Binary Fading IC with global delayed CSIT, $\mathcal{C}\left( \text{V.}8 \right)$, is the set of all rate-tuples $\left( R_1, R_2 \right)$ satisfying
\begin{equation}
\label{eq:DelayedCSIregion}
\left\{ \begin{array}{ll}
\vspace{1mm} 0 \leq R_i \leq p, &  i = 1,2, \\
R_i + \left( 1 + q \right) R_{\bar{i}} \leq p \left( 1 + q \right)^2, & i = 1,2.
\end{array} \right.
\end{equation}

\begin{figure}[ht]
\centering
\includegraphics[height = 4.5cm]{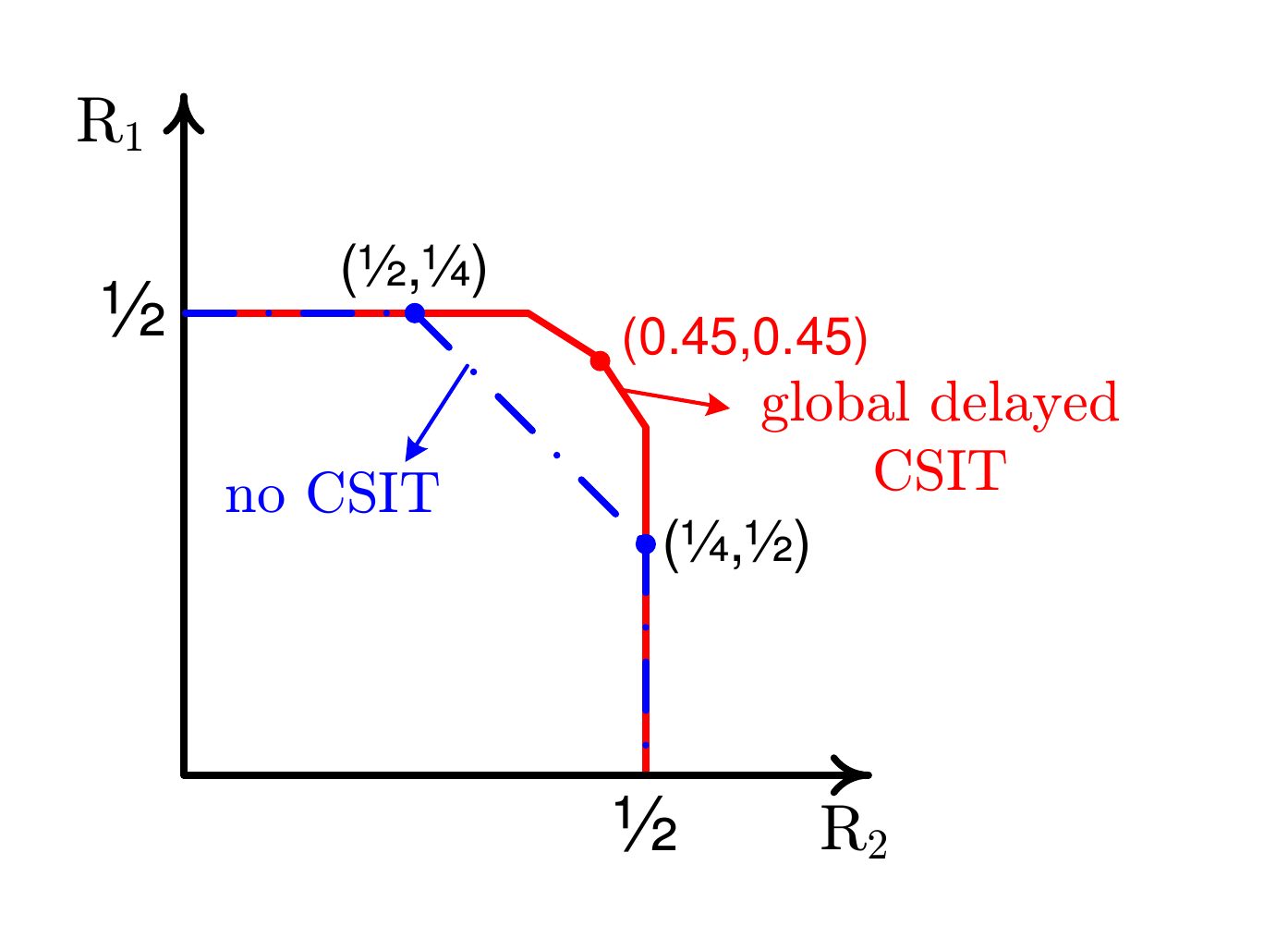}
\caption{\it Capacity Region of the Two-user Binary Fading Interference Channel with no and global delayed CSIT and $p = 0.5$.\label{Fig:NovsGlobal}}
\end{figure}

These benchmarks are depicted in Fig.~\ref{Fig:NovsGlobal} for $p = 0.5$. We note that $\mathcal{C}\left( \text{V.}0 \right)$ is a strict subset of $\mathcal{C}\left( \text{V.}8 \right)$, and we are interested in understanding the impact of local delayed CSIT on the capacity region.

\subsection{Main Results}

Theorem~\ref{THM:NoGain} highlights the cases where no performance gain over the no CSIT assumption is feasible. Then in Theorem~\ref{THM:FullGain}, we show that the performance gain of global delayed CSIT (View V.8) can be obtained with Views V.2, V.5, and V.6. Note that the achievable region for View V.2 is a subset of Views V.5 and V.6 due to the hierarchical structure shown in Fig.~\ref{Fig:Hierarchy}.

\begin{theorem}
\label{THM:NoGain}
For the two-user binary fading interference channel with local delayed CSIT of Views V.1, V.3, and V.4, the capacity region coincides with the capacity region of no CSIT (View V.0), \emph{i.e.}
\begin{align}
\mathcal{C}\left( \text{V.}j \right) \subseteq \mathcal{C}\left( \text{V.}0 \right) \text{~for~} \qquad j=1,3,4.
\end{align}
\end{theorem}

The key in proving Theorem~\ref{THM:NoGain} is to derive an outer-bound the matches that of the no CSIT assumption. In doing that, we use the intuition given by Sato~\cite{sato1977two}: the capacity region of all interference channels that have the same marginal distributions is the same. We take this intuition and impose a certain spatial correlation among channel gains such that the marginal distributions remain unchanged. Then we provide an outer-bound on the capacity region of the channel with correlation and we show that it can be achieved with no CSIT.

\begin{remark}
In this paper, we only focus on the impact of local {\it delayed} CSIT on the capacity region of the two-user erasure ICs. Interestingly, the proof of Theorem~\ref{THM:NoGain} holds even when channels are learned instantaneously.
\end{remark}

\begin{theorem}
\label{THM:FullGain}
For the two-user binary fading interference channel with local delayed CSIT of Views V.2, V.5, and V.6, the capacity region coincides with the capacity region of global delayed CSIT (V.8), \emph{i.e.}
\begin{align}
\mathcal{C}\left( \text{V.}8 \right) \subseteq \mathcal{C}\left( \text{V.}j \right) \text{~for~} \qquad j=2,5,6.
\end{align}
\end{theorem}

For View V.2, we need to provide an achievability strategy that achieves the same performance as V.8. Then the result for View V.5 and V.6 follows. The capacity region $\mathcal{C}\left( \text{V.}8 \right)$ was established in~\cite{AlirezaBFICDelayed} and a novel transmission strategy was introduced. The strategy in~\cite{AlirezaBFICDelayed} is carried on over several phases. Each channel realization creates multiple coding opportunities which can be exploited in the following phases. To achieve the capacity region, an efficient arrangement of combination, concatenation, and merging of the opportunities is needed. Our result for View V.2 provides a better intuition and a deeper understanding of such opportunities and reveals redundancies in the prior work.

\begin{remark}
The capacity region under View V.7 remains open. While it seems the coding opportunities cannot be detected with such channel state information, the lack of an outer-bound does not allow us to solve the problem for this case. In a sense, each transmitter has ``too much'' knowledge eliminating the choices for correlation that were needed to obtain the result in Theorem~\ref{THM:NoGain}. We discuss this case in more details in Section~\ref{Section:Discussion}.
\end{remark}

In the remaining of the paper, we provide the proof of Theorem~\ref{THM:NoGain} and Theorem~\ref{THM:FullGain}. In Section~\ref{Section:Discussion}, we discuss the challenges for View V.7 and provide some connections to the $k$-user setting ($k > 2$).


\section{Proof of Theorem~\ref{THM:NoGain}: Choosing Spatial Correlations}
\label{Section:NoGain}

In this section we provide the proof of Theorem~\ref{THM:NoGain}. In particular, we demonstrate that with local delayed CSIT according to Views V.1, V.3, and V.4 the capacity region is the same as View V.0 (no CSIT) as given in (\ref{eq:RegionNoCSIT}). Since the knowledge in View V.1 is a subset of other cases, we do not need to provide a separate proof for this case. Thus, we only need to provide the proof for Views V.3 and V.4. 

Here, we introduce a certain spatial correlation among channel gains such that the marginal distributions remain unchanged. Then, we provide an outer-bound on the capacity region of the channel with correlation that matches the no CSIT region of (\ref{eq:RegionNoCSIT}). The following facts will help us throughout the proof. For any choice of local delayed CSIT, we have
\begin{align}
\label{eq:ConditionalIndependence}
\Pr \left( X_1^n, X_2^n | G^n \right) = \Pr \left( X_1^n | G^n \right) \Pr \left( X_2^n | G^n \right).
\end{align}
Moreover, if $\mathcal{S}_{{\sf Tx}_1} \cap \mathcal{S}_{{\sf Tx}_2} = \emptyset$, then we have
\begin{align}
\label{eq:Independence}
\Pr \left( X_1^n, X_2^n \right) = \Pr \left( X_1^n \right) \Pr \left( X_2^n \right).
\end{align}
Derivation of (\ref{eq:ConditionalIndependence}) and (\ref{eq:Independence}) is a straightforward exercise and is omitted here.

In~\cite{AlirezaNoCSIT}, it was shown that under local CSI at the receivers (CSIR):
\begin{align}
\mathcal{C}\left( \text{V.}1 \right) \subseteq \mathcal{C}\left( \text{V.}0 \right).
\end{align}
In fact, the proof relied heavily on the assumption of local CSIR and cannot be extended to our setting where the receivers know the {\it entire} CSI instantaneously.

\subsection{Proof for View V.4}

We know that the error probabilities are solely a function of marginal distributions at the receivers. Thus, as long as the marginal distributions remain the same, the capacity region remains the same. Consider the two-user erasure IC with local delayed CSIT according to View V.4. We have
\begin{align} 
\mathcal{S}_{{\sf Tx}_1} = \{ \left( 1, 1 \right), \left( 2, 1 \right) \} \quad \text{and} \quad \mathcal{S}_{{\sf Tx}_2} = \{ \left( 1, 2 \right), \left( 2, 2 \right) \}.
\end{align}
Thus writing the marginal distribution at receiver ${\sf Rx}_1$, we get
\begin{align}
\label{eq:MarginalView4Rx1}
\Pr & \left( Y_1^n, G^n | X_1^n, X_2^n \right) \nonumber \\
& \overset{(a)}= \Pr \left( G^n | X_1^n, X_2^n \right) \Pr \left( Y_1^n |  X_1^n, X_2^n, G^n \right) \nonumber \\
& = \frac{\Pr \left( G^n, X_1^n, X_2^n \right)}{\Pr \left( X_1^n, X_2^n \right)} \Pr \left( Y_1^n |  X_1^n, X_2^n, G^n \right) \nonumber \\
& \overset{(b)}= \frac{\Pr \left( G^n \right) \Pr \left( X_1^n | G^n \right) \Pr \left( X_2^n | G^n \right)}{\Pr \left( X_1^n \right) \Pr \left( X_2^n \right)} \Pr \left( Y_1^n |  X_1^n, X_2^n, G^n \right) \nonumber \\
& \overset{(c)}= \frac{\Pr \left( G^n \right) \Pr \left( X_1^n | G_{11}^n, G_{21}^n \right) \Pr \left( X_2^n | G_{12}^n, G_{22}^n \right)}{\Pr \left( X_1^n \right) \Pr \left( X_2^n \right)} \nonumber \\
& ~\times \Pr \left( Y_1^n |  X_1^n, X_2^n, G^n \right) \nonumber \\
& = \left[ \frac{\Pr \left( G_{11}^n, G_{21}^n \right) \Pr \left( X_1^n | G_{11}^n, G_{21}^n \right)}{\Pr \left( X_1^n \right)} \right] \nonumber \\
& ~\times \left[ \frac{\Pr \left( G_{12}^n, G_{22}^n \right) \Pr \left( X_2^n | G_{12}^n, G_{22}^n \right)}{\Pr \left( X_2^n \right)} \right] \nonumber \\
& ~\times \mathbf{1}_{\left\{ Y_1^n = G_{11}^n X_1^n \oplus G_{21}^n X_2^n \right\}},
\end{align}
where $(a)$ follows from the chain rule; $(b)$ follows from the chain rule and the fact that $X_1^n$ and $X_2^n$ are independent in View V.4 as given in (\ref{eq:Independence}); and $(c)$ holds since $X_i^n$ is independent of $G_{i\bar{i}}^n$ and $G_{\bar{i}\bar{i}}^n$, $i=1,2$. Similarly, we can write the marginal distribution at receiver ${\sf Rx}_2$.
\begin{align}
\label{eq:MarginalView4Rx2}
& \Pr\left( Y_2^n, G^n | X_1^n, X_2^n \right) = \left[ \frac{\Pr \left( G_{11}^n, G_{21}^n \right) \Pr \left( X_1^n | G_{11}^n, G_{21}^n \right)}{\Pr \left( X_1^n \right)} \right] \nonumber \\
& \times \left[ \frac{\Pr \left( G_{12}^n, G_{22}^n \right) \Pr \left( X_2^n | G_{12}^n, G_{22}^n \right)}{\Pr \left( X_2^n \right)} \right] \mathbf{1}_{\left\{ Y_2^n = G_{12}^n X_1^n \oplus G_{22}^n X_2^n \right\}}.
\end{align}

We conclude that as long as the joint distributions 
\begin{align}
\Pr \left( G_{11}^n, G_{21}^n \right) \quad \text{and} \quad \Pr \left( G_{12}^n, G_{22}^n \right)
\end{align}
remain the same, the marginal distributions remain unchanged. This immediately implies that the capacity region $\mathcal{C}\left( \text{V.}4 \right)$ coincides with the capacity region of any channel that has same marginal distributions  $\Pr \left( G_{11}^n, G_{21}^n \right)$ and $\Pr \left( G_{12}^n, G_{22}^n \right)$. We use this result to create a channel with a specific correlation that helps us deriving the desired result.

Consider a binary fading interference channel similar to the channel described in Section~\ref{Section:Problem}, but where channel gains have certain spatial correlation. We distinguish the RVs in this channel using $\left( \tilde{.} \right)$ notation (\emph{e.g.}, $\tilde{X}_1[t]$). The input-output relation of this channel at time instant $t$ is given by
\begin{equation} 
\label{Eq:modifiedView4}
\tilde{Y}_i[t] = \tilde{G}_{ii}[t] \tilde{X}_i[t] \oplus \tilde{G}_{\bar{i}i}[t] \tilde{X}_{\bar{i}}[t], \quad i = 1, 2.
\end{equation}

We assume that the channel gains are distributed independently over time. However, we have
\begin{align}
\label{eq:constraint}
\tilde{G}_{ii}[t] = \tilde{G}_{i\bar{i}}[t] \quad i=1,2.
\end{align}
In other words, the channel gains corresponding to incoming links at each receiver are still independent but the outgoing links at each transmitter are correlated. We know that the capacity region of this channel coincides with $\mathcal{C}\left( \text{V.}4 \right)$. We have included a more detailed discussion in Appendix~\ref{Appendix:NewAppendix}. 

Suppose rate-tuple $\left( \tilde{R}_1, \tilde{R}_2 \right)$ is achievable. The derivation of individual bounds for this channel
\begin{align}
\label{eq:individualbound}
\tilde{R}_i \leq p, \qquad i=1,2,
\end{align}
is a straight forward exercise and omitted here. For the sum-rate bound, we have
\begin{align}
n & \left( \tilde{R}_1 + \tilde{R}_2 - \epsilon_n \right) \overset{(a)}\leq I\left( \tilde{W}_1; \tilde{Y}_1^n | \tilde{W}_2, \tilde{G}^n \right) + I\left( \tilde{W}_2; \tilde{Y}_2^n | \tilde{G}^n \right) \nonumber \\
& = H\left( \tilde{Y}_1^n | \tilde{W}_2, \tilde{G}^n \right) - \underbrace{H\left( \tilde{Y}_1^n | \tilde{W}_1, \tilde{W}_2, \tilde{G}^n \right)}_{=~0}\nonumber \\
&~+ H\left( \tilde{Y}_2^n | \tilde{G}^n \right) - H\left( \tilde{Y}_2^n | \tilde{W}_2, \tilde{G}^n \right) \nonumber \\
& = H\left( \tilde{Y}_2^n | \tilde{G}^n \right) + H\left( \tilde{G}_{11}^n \tilde{X}_1^n | \tilde{X}_2^n, \tilde{W}_2, \tilde{G}^n \right) \nonumber \\
& ~- H\left( \tilde{G}_{12}^n \tilde{X}_1^n | \tilde{X}_2^n, \tilde{W}_2, \tilde{G}^n \right) \nonumber \\
& = H\left( \tilde{Y}_2^n | \tilde{G}^n \right) + H\left( \tilde{G}_{11}^n \tilde{X}_1^n | \tilde{G}^n \right) - H\left( \tilde{G}_{12}^n \tilde{X}_1^n | \tilde{G}^n \right) \nonumber \\
& \overset{(b)}= H\left( \tilde{Y}_2^n | \tilde{G}^n \right) \leq \left( 1 - q^2 \right)n,
\end{align}
where $\epsilon_n \rightarrow 0$ as $n \rightarrow \infty$; $(a)$ follows from Fano's inequality and the fact that messages and channel gains are mutually independent; $(b)$ holds since according to (\ref{Eq:modifiedView4}), we have 
\begin{align}
H\left( \tilde{G}_{11}^n \tilde{X}_1^n | \tilde{G}^n \right) = H\left( \tilde{G}_{12}^n \tilde{X}_1^n | \tilde{G}^n \right).
\end{align}
Dividing both sides by $n$ and letting $n \rightarrow \infty$, we obtain
\begin{align}
\label{eq:sumratebound}
\tilde{R}_1 + \tilde{R}_2 \leq 1-q^2.
\end{align}

Note that the region described by (\ref{eq:individualbound}) and (\ref{eq:sumratebound}) matches the no CSIT region of (\ref{eq:RegionNoCSIT}). This completes the converse proof for View V.4 since $\mathcal{C}\left( \text{V.}4 \right)$ is included in the region described by (\ref{eq:individualbound}) and (\ref{eq:sumratebound}).

\begin{remark}
The proof does not rely on fact that the CSI is obtained with delay. In fact, if we assume instantaneous local CSIT, the proof still holds. However in this paper, we only focus on local {\it delayed} CSIT for which we present achievability strategy as well. 
\end{remark} 

\subsection{Proof for View V.3}

The proof is very similar to that of View V.4. Under local delayed CSIT of View V.3, we have
\begin{align} 
\mathcal{S}_{{\sf Tx}_1} = \{ \left( 1, 1 \right), \left( 2, 2 \right) \} \quad \text{and} \quad \mathcal{S}_{{\sf Tx}_2} = \{ \left( 1, 1 \right), \left( 2, 2 \right) \}.
\end{align}
Thus writing the marginal distribution for receiver ${\sf Rx}_1$, we get
\begin{align}
\label{eq:MarginalView3Rx1}
& \Pr \left( Y_1^n, G^n | X_1^n, X_2^n \right) \nonumber \\
& = \Pr \left( X_1^n,X_1^n \right)^{-1}\left[ \Pr \left( G_{12}^n, G_{21}^n \right) \Pr \left( X_1^n | G_{11}^n, G_{22}^n \right) \right] \nonumber \\
& \times \left[ \Pr \left( G_{11}^n, G_{22}^n \right) \Pr \left( X_2^n | G_{11}^n, G_{22}^n \right) \right]  \mathbf{1}_{\left\{ Y_1^n = G_{11}^n X_1^n \oplus G_{21}^n X_2^n \right\}},
\end{align}
Similarly, we can write the marginal distribution for receiver ${\sf Rx}_2$.
\begin{align}
\label{eq:MarginalView3Rx2}
& \Pr \left( Y_2^n, G^n | X_1^n, X_2^n \right) \nonumber \\
& = \Pr \left( X_1^n,X_1^n \right)^{-1} \left[ \Pr \left( G_{12}^n, G_{21}^n \right) \Pr \left( X_1^n | G_{11}^n, G_{22}^n \right) \right] \nonumber \\
& \times \left[ \Pr \left( G_{11}^n, G_{22}^n \right) \Pr \left( X_2^n | G_{11}^n, G_{22}^n \right) \right] \mathbf{1}_{\left\{ Y_2^n = G_{12}^n X_1^n \oplus G_{22}^n X_2^n \right\}}.
\end{align}

We conclude that as long as the joint distributions 
\begin{align}
\Pr \left( G_{11}^n, G_{21}^n \right), \quad \Pr \left( G_{12}^n, G_{22}^n \right), \quad \text{and} \quad \Pr \left( G_{11}^n, G_{22}^n \right),
\end{align}
remain the same, the capacity region remains the same.  We note that the same correlation introduced in (\ref{Eq:modifiedView4}) can be applied here. Thus, the rest of the proof is identical to the previous subsection.


\section{Proof of Theorem~\ref{THM:FullGain}: Opportunistic Retransmissions Based on Local Delayed CSIT}
\label{Section:FullGain}

In this section, we provide an achievability strategy that solely relies on what the transmitters know in View V.2, \emph{i.e.} the outdated knowledge of the channel gains associated with the outgoing links at each transmitter. We show that the capacity region with global delayed CSIT can be achieved with this local delayed knowledge. This result is surprising since the transmission strategy in prior work~\cite{AlirezaBFICDelayed} for the case of global delayed CSIT relies heavily on the delayed knowledge of the entire channel state information at each transmitter. In fact, the transmission strategy of~\cite{AlirezaBFICDelayed} cannot be applied to the case where transmitter learn the CSI locally.

We note that the channel knowledge of View V.2 is a subset of View V.5 or View V.6; thus there is no need to prove the result separately for those cases. Prior to providing the achievability proof, we discuss some techniques and coding opportunities that we utilize later in this section.

\subsection{Techniques and Coding Opportunities}
\label{Section:Opportunities}

In this subsection, we describe the coding opportunities that arise from the delayed knowledge of the channel state information. We first assume that nodes have global delayed CSIT to clearly describe the opportunities. Then we discuss the challenges that stem from the locality of the channel state information at the transmitters. Next, we demonstrate how to overcome these challenges. Finally in the following subsection, we describe the transmission strategy in detail. We categorize the opportunities into three groups as follows.

\begin{figure}[ht]
\centering
\subfigure[]{\includegraphics[height = 3 cm]{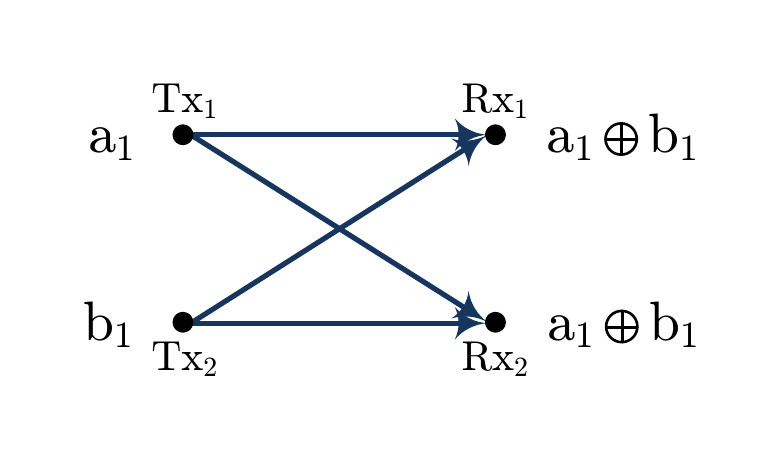}}
\hspace{0.1 in}
\subfigure[]{\includegraphics[height = 3 cm]{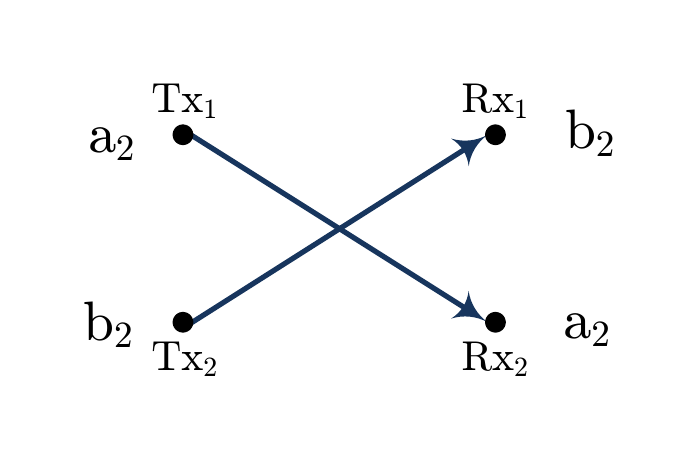}}
\caption{\it Providing $a_1 \oplus a_2$ available at ${\sf Tx}_1$ and $b_1 \oplus b_2$ available at ${\sf Tx}_2$ to both receivers is sufficient to decode the bits.\label{Fig:Opp1}}
\end{figure}

[{Pairing Across Realizations Type-I}] Suppose at a time instant, each one of the transmitters simultaneously sends one data bit. The bits of ${\sf Tx}_1$ and ${\sf Tx}_2$ are denoted by $a_1$ and $b_1$ respectively. Assume that the channel realization was according to Fig.~\ref{Fig:Opp1}(a). In another time instant, each one of the transmitters sends one data bit, say $a_2$ and $b_2$ respectively. Assume that the channel realization was according to Fig.~\ref{Fig:Opp1}(b). Now, we observe that providing $a_1 \oplus a_2$ and $b_1 \oplus b_2$ to both receivers is sufficient to decode the bits. For instance if ${\sf Rx}_1$ is provided with $a_1 \oplus a_2$ and $b_1 \oplus b_2$, then it will use $b_2$ to decode $b_1$, from which it can obtain $a_1$, and finally using $a_1$ and $a_1 \oplus a_2$, it can decode $a_2$. The linear combinations $a_1 \oplus a_2$ and $b_1 \oplus b_2$ that are available at transmitters one and two respectively, can be thought of as bits of ``common interest.''

\begin{figure}[ht]
\centering
\subfigure[]{\includegraphics[height = 3 cm]{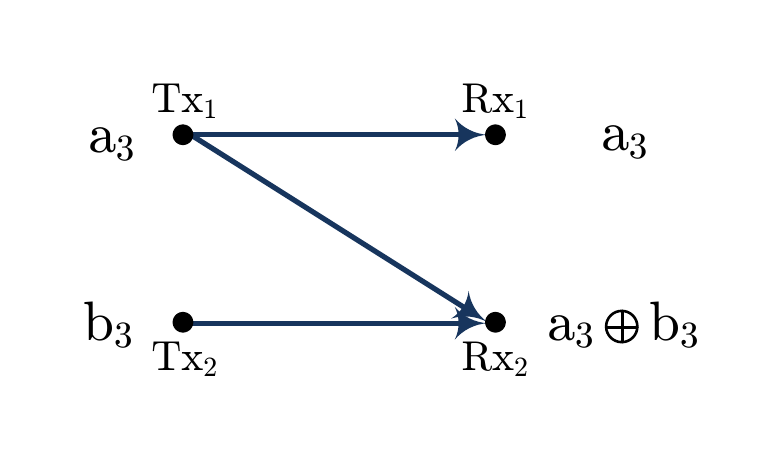}}
\hspace{0.1 in}
\subfigure[]{\includegraphics[height = 3 cm]{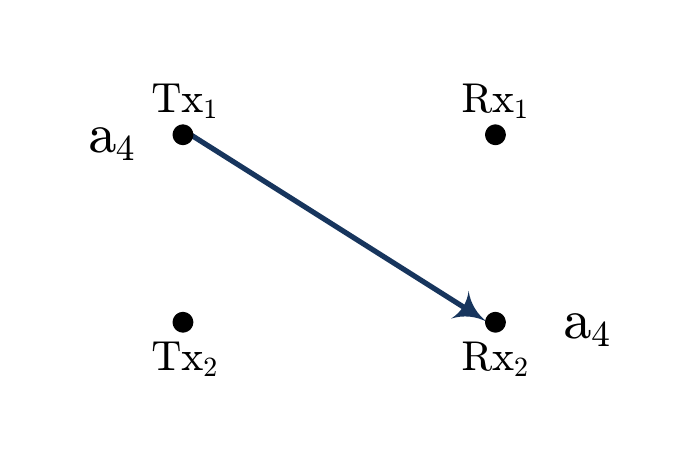}}
\caption{\it The combination $a_3 \oplus a_4$ available at ${\sf Tx}_1$ is of interest of both receivers.\label{Fig:Opp2}}
\end{figure}

[{Pairing Across Realizations Type-II}] Suppose the scenario depicted in Fig.~\ref{Fig:Opp2} is realized. Now, we observe that providing $a_3 \oplus a_4$ available at transmitter ${\sf Tx}_1$ to both receivers is useful. For instance if ${\sf Rx}_2$ is provided with $a_3 \oplus a_4$, it will use $a_4$ to decode $a_3$, from which it can obtain $b_3$. It is easy to visualize the similar opportunity for transmitter ${\sf Tx}_2$.

\begin{figure}[ht]
\centering
\subfigure[]{\includegraphics[height = 3 cm]{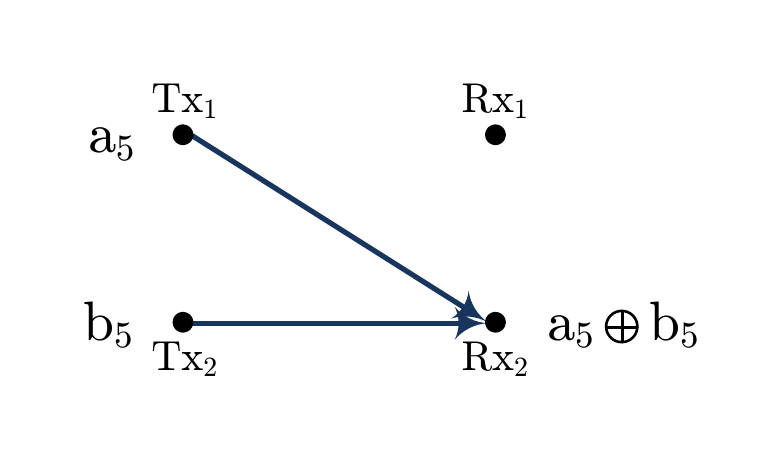}}
\hspace{0.1 in}
\subfigure[]{\includegraphics[height = 3 cm]{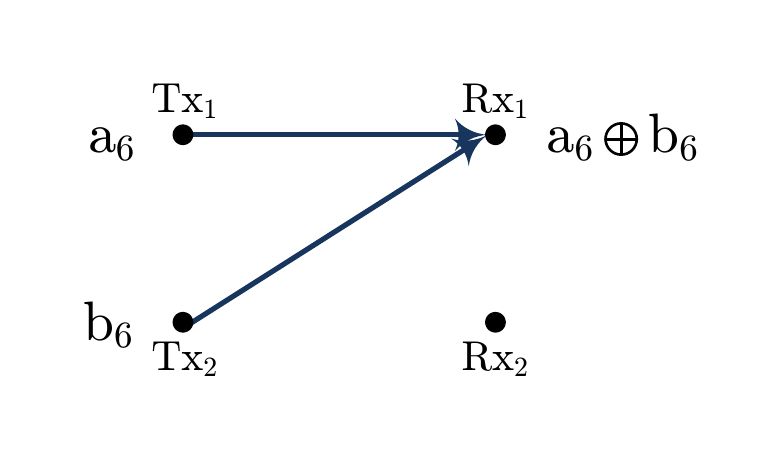}}
\caption{\it In each of the channel realizations, one bit $a_5$ becomes of interest of both receivers. In (a) bit $a_5$ is a useful bit for both receivers, while in (b) bit $b_6$ is a useful bit for both receivers.\label{Fig:Opp3}}
\end{figure}

[{Pairing Across Realizations Type-III}] Suppose the scenario depicted in Fig.~\ref{Fig:Opp3}(a) has occurred. It is easy to see that bit $a_5$ is a useful bit for {\it both} receivers after this point. A similar situation is depicted in Fig.~\ref{Fig:Opp3}(b) where bit $b_6$ at transmitter ${\sf Tx}_2$ becomes a bit of common interest.

\begin{table*}[t]
\caption{The four possible configurations that can be identified by transmitter ${\sf Tx}_1$. The bit transmitted by ${\sf Tx}_1$ is denoted ``$a$.'' Depending on the identified configuration, the status of the transmitted bit is updated to a queue defined in Section~\ref{Section:TStrategy}.}
\centering
\begin{tabular}{| c | c | c | c | c | c |}
\hline
case ID		 & channel realization    & state transition  & case ID		 & channel realization    & state transition \\
					 & at time instant $n$    &                   & 					 & at time instant $n$    &                  \\ [0.5ex]
\hline

\raisebox{30pt}{$1$}    &    \includegraphics[height = 2.5cm]{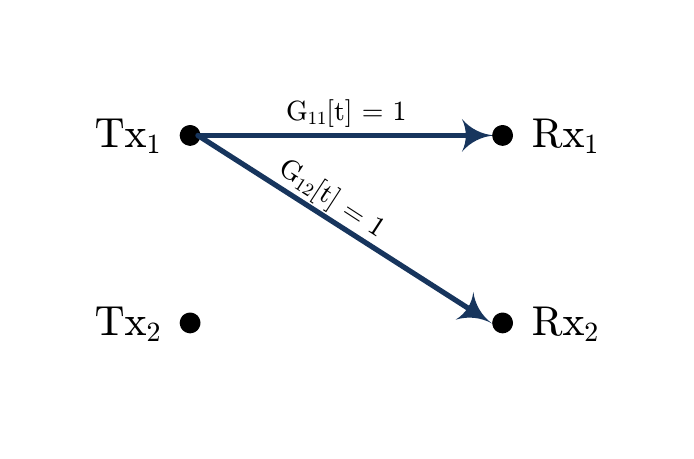}     &  \raisebox{30pt}{$a \rightarrow Q_{1,1}$}  &  \raisebox{18pt}{$3$}    &    \includegraphics[height = 2.5cm]{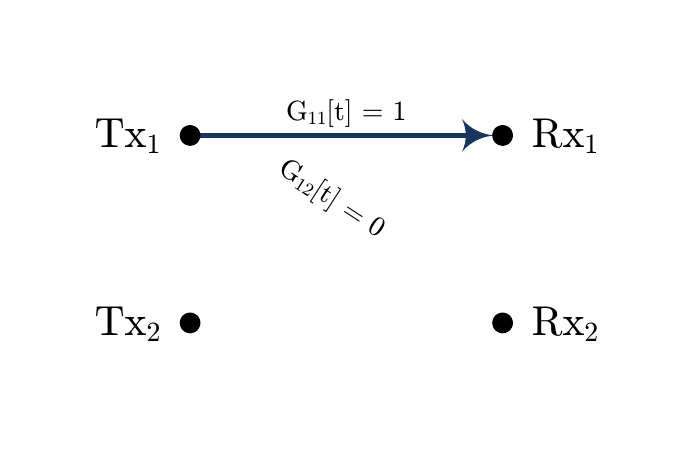}     &   \raisebox{18pt}{$a \rightarrow Q_{1 \rightarrow F}$} \\

\hline

\raisebox{30pt}{$2$}    &    \includegraphics[height = 2.5cm]{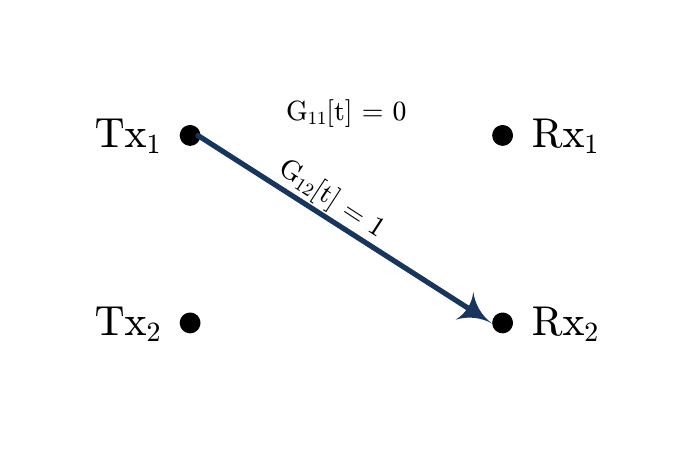}     &  \raisebox{30pt}{$a \rightarrow Q_{1,2}$} &  \raisebox{18pt}{$4$}    &    \includegraphics[height = 2.5cm]{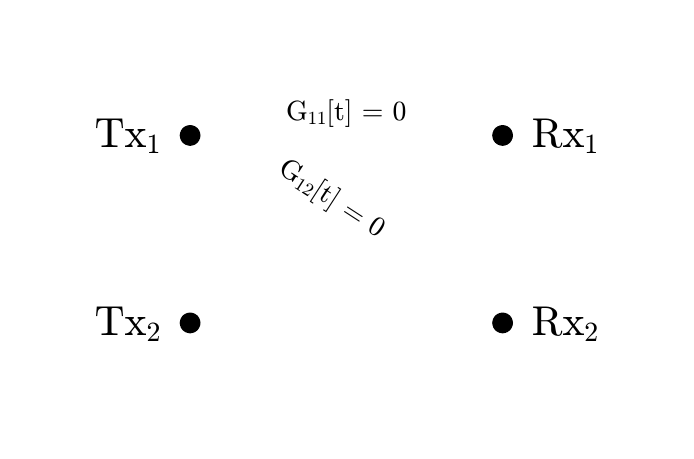}     &   \raisebox{18pt}{$a \rightarrow Q_{1 \rightarrow 1 }$} \\

\hline

\end{tabular}
\label{Table:View2}
\end{table*}

\begin{figure}[ht]
\centering
\subfigure[]{\includegraphics[height = 3 cm]{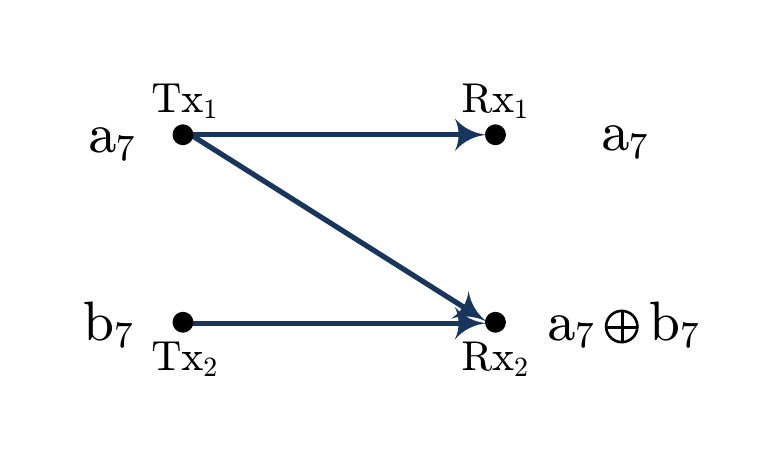}}
\hspace{0.1 in}
\subfigure[]{\includegraphics[height = 3 cm]{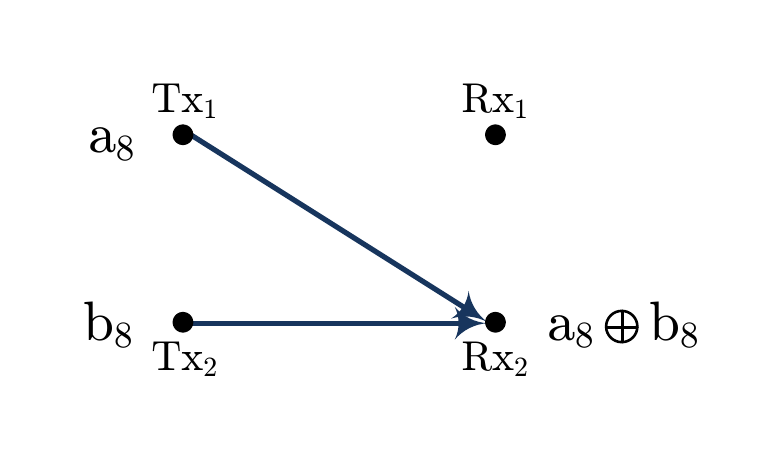}}
\caption{\it From transmitter ${\sf Tx}_1$'s point of view, this scenario is identical to the one depicted in Fig.~\ref{Fig:Opp1} and Fig.~\ref{Fig:Opp2}. However, $a_7 \oplus a_8$ is no longer useful for ${\sf Rx}_2$.\label{Fig:OppBad}}
\end{figure}

{\bf Challenges with local delayed CSIT:} Given local delayed CSIT of View V.2, each transmitter can identify a total of $4$ possible configurations as summarized in Table~\ref{Table:View2} for ${\sf Tx}_1$ (the configurations known to ${\sf Tx}_2$ are simply derived by interchanging user IDs). With the limited knowledge available at each transmitter, the aforementioned opportunities may not be detected properly. For instance consider the scenario depicted in Fig.~\ref{Fig:OppBad}. From transmitter ${\sf Tx}_1$'s point of view, this scenario is identical to the one depicted in Fig.~\ref{Fig:Opp1} and Fig.~\ref{Fig:Opp2}. However, providing $a_5 \oplus a_6$ to ${\sf Rx}_2$ is not useful anymore. The proposed scheme of~\cite{AlirezaBFICDelayed} cannot overcome this challenge and fails with local knowledge as a result. Therefore, one must be careful on how to identify the opportunities and how to exploit them for future communications.

\begin{remark}
It is important to keep in mind that while transmitters have only local knowledge of the CSI, receivers have global knowledge. This enables receivers to figure out future actions taken by the transmitters based on their past observations of the channel realizations.
\end{remark}

{\bf Overcoming the challenges with local delayed CSIT:} Consider the example demonstrated in Fig.~\ref{Fig:OppBad}. Transmitter ${\sf Tx}_1$ can overcome the challenge described above by relying on the fact that statistically a fraction $q$ of the bits that at the time of transmission faced $G_{12}[t] = 1$, are already known to ${\sf Rx}_2$ (since with probability $q$ we have $G_{22}[t] = 0$). Keeping this fact in mind, transmitter ${\sf Tx}_1$ will create enough linearly independent combinations of the bits such that receiver ${\sf Rx}_2$ can recover the required bits. The global knowledge at the receivers is essential for them to know which bits are going to be retransmitted and in what order. This technique is described in more detail in the following subsection.

We also note that out of the four configurations in Table~\ref{Table:View2}, in configurations 3 and 4 the future task of ${\sf Tx}_1$ is easy. Suppose at time $t$, transmitter ${\sf Tx}_1$ sends one data bit. Later, using local delayed CSIT, ${\sf Tx}_1$ figures out that $G_{11}[t] = 1$ and $G_{12}[t] = 0$. In this case, we say that the bit is delivered and if interference was created at ${\sf Rx}_1$ then it would be the responsibility of ${\sf Tx}_2$ to resolve it in future. If $G_{11}[t]$ and $G_{12}[t]$ were both equal to $0$, then the bit must be retransmitted.

We show that with only local delayed CSIT of View V.2, we can achieve $\mathcal{C}\left( \text{V.}8 \right)$. 

\subsection{Transmission Strategy}
\label{Section:TStrategy}

In this Section, we focus on the achievability strategy for $p = 0.5$. This would simplify the transmission strategy and allows us to focus on the key issue of local delayed CSIT. 

\begin{figure}[ht]
\centering
\includegraphics[height = 4.5cm]{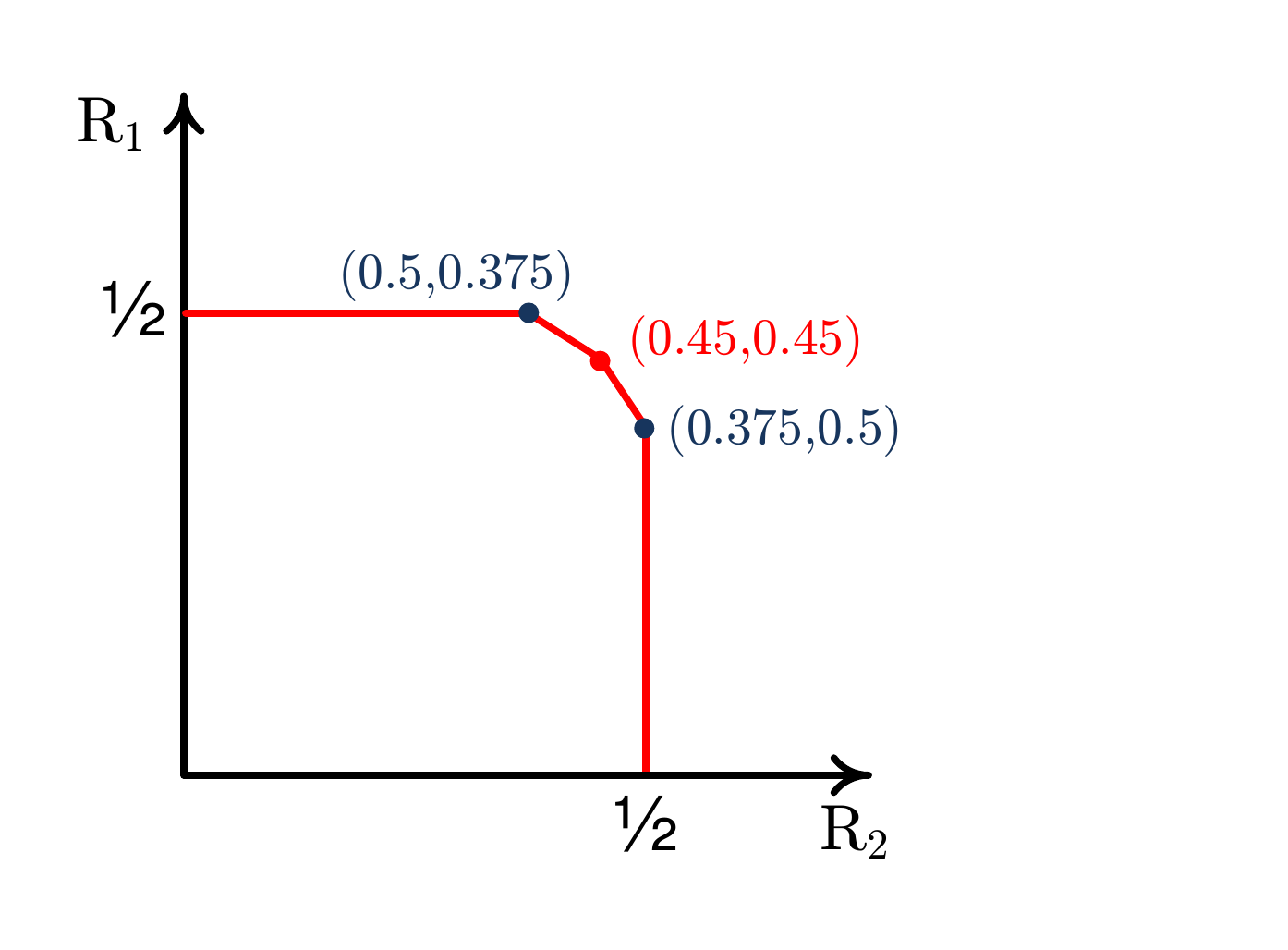}
\caption{\it To show that with only local delayed CSIT of View V.2, we can achieve $\mathcal{C}\left( \text{V.}8 \right)$, it suffices to prove the achievability for corner points $\left( 0.45, 0.45 \right)$ and $\left( 0.375, 0.5 \right)$.\label{Fig:RegionFull}}
\end{figure}

We have again depicted the capacity region with global delayed CSIT $\mathcal{C}\left( \text{V.}8 \right)$ in Fig.~\ref{Fig:RegionFull} for $p = 0.5$. We show that with only local delayed CSIT of View V.2, we can achieve $\mathcal{C}\left( \text{V.}8 \right)$. To do so, it suffices to prove achievability for the corner points $\left( 0.45, 0.45 \right)$ and $\left( 0.375, 0.5 \right)$.

We focus on the corner point $\left( R_1, R_2 \right) = \left( 0.45, 0.45 \right)$. The achievability strategy for the other corner points is presented in Appendix~\ref{Appendix:Corner}. Then in Appendix~\ref{Appendix:GeneralP}, we describe the changes needed in the transmission strategy when considering $0 \leq p \leq 1$.  

Suppose each transmitter wishes to communicate $m$ bits to its intended receiver. We show that this task can be accomplished (with vanishing error probability as $m \rightarrow \infty$) in
\begin{align}
\frac{20}{9} m + \frac{35}{3} m^{\frac{2}{3}}
\end{align}
time instants. This immediately implies the achievability for the corner point $\left( R_1, R_2 \right) = \left( 0.45, 0.45 \right)$. Our transmission strategy comprises two phases as described below. 

\noindent {\bf Phase 1}: At the beginning of the communication block, we assume that the $m$ bits at ${\sf Tx}_i$ are in queue $Q_{i \rightarrow i}$ (the initial state of the bits), $i=1,2$. At each time instant $t$, ${\sf Tx}_i$ sends out a bit from $Q_{i \rightarrow i}$, and this bit will either stay in the initial queue or transition to one of the queues listed in Table~\ref{Table:View2}. If at time instant $t$, $Q_{i \rightarrow i}$ is empty, then ${\sf Tx}_i$, $i=1,2$, remains silent until the end of Phase 1. 
\begin{enumerate}
\item [(A)] $Q_{i \rightarrow F}$: The bits for which no retransmission is required and thus we consider delivered;

\item [(B)] $Q_{i, 1}$: The bits for which at the time of communication, all channel gains known to ${\sf Tx}_i$ with unit delay were equal to $1$;

\item [(C)] $Q_{i, 2}$: The bits for which at the time of communication, we have $G_{ii}[t] = 0$ and $G_{i\bar{i}}[t] = 1$.
\end{enumerate}

Each transmitter can identify a total of $4$ possible configurations as summarized in Table~\ref{Table:View2} for ${\sf Tx}_1$. Phase~$1$ continues for 
\begin{align}
\frac{4}{3} m + m^{\frac{2}{3}}
\end{align}
time instants, and if at the end of this phase, either of the queues $Q_{i \rightarrow i}$ is not empty, we declare error type-I and halt the transmission (we assume $m$ is chosen such that $m^{\frac{2}{3}} \in \mathbb{Z}$). We assume that the queues are column vectors and bits are placed according to the order they join the queue.

Assuming that the transmission is not halted, let $N_{i,1}$ and $N_{i, 2}$ denote the number of bits in queues $Q_{i,1}$ and $Q_{i,2}$ respectively at the end of the transitions, $i=1,2$. The transmission strategy will be halted and error type-II occurs, if any of the following events happens.
\begin{align}
\label{eq:errortypeII}
& N_{i,1} > \mathbb{E}[N_{i,1}] + 2 m^{\frac{2}{3}} \overset{\triangle}= n_{i,1}, \quad i=1,2; \nonumber \\
& N_{i,2} > \mathbb{E}[N_{i,2}] + 2 m^{\frac{2}{3}} \overset{\triangle}= n_{i,2}, \quad i=1,2.
\end{align}

From basic probability, we have
\begin{align}
\label{eq:expectedvalues}
\mathbb{E}[N_{i,1}] = \mathbb{E}[N_{i,2}] = \frac{m}{3},
\end{align}
thus we get
\begin{align}
 n_{i,1} =  n_{i,2} = \frac{m}{3} + 2 m^{\frac{2}{3}}.
\end{align}
At the end of Phase $1$, we add $0$'s (if necessary) in order to make queues $Q_{i,1}$ and $Q_{i,2}$ of size equal to $n_{i,1}$ and $n_{i,2}$ respectively as given above, $i=1,2$.

Moreover since channel gains are distributed independently, statistically half of the bits in $Q_{i,1}$ and half of the bits in $Q_{i,2}$ are known to ${\sf Rx}_{\bar{i}}$, $i=1,2$. Denote the number of bits in $Q_{i,j}$ known to ${\sf Rx}_{\bar{i}}$ by
\begin{align}
N_{i,j|{\sf Rx}_{\bar{i}}}, \qquad i,j \in \{ 1, 2 \}.
\end{align}
At the end of communication, if we have
\begin{align}
\label{Eq:KnownBits}
N_{i,j|{\sf Rx}_{\bar{i}}} < \frac{1}{2} n_{i,j} - m^{\frac{2}{3}}, \qquad i,j \in \{ 1, 2 \},
\end{align}
we declare error type-III. Note that transmitters cannot detect error type-III, but receivers have sufficient information to do so. 

Furthermore using the Bernstein inequality, we can show that the probability of errors of types I, II, and III decreases exponentially with $m$. For the rest of this subsection, we assume that Phase~1 is completed and no error has occurred.

Transmitter ${\sf Tx}_i$ creates two matrices $\mathbf{C}_{i,1}$ and $\mathbf{C}_{i,2}$, $i=1,2$, of size $\left( \frac{m}{3} + 4 m^{\frac{2}{3}} \right) \times \left( \frac{m}{3} + 2 m^{\frac{2}{3}} \right)$ each, where entries to each matrix are drawn from i.i.d. $\mathcal{B}(0.5)$ distribution. We assume that these matrices are generated prior to communication and are shared with receivers. Transmitter ${\sf Tx}_i$ does not need to know $\mathbf{C}_{\bar{i},1}$ or $\mathbf{C}_{\bar{i},2}$, $i=1,2$. Note that as $m \rightarrow \infty$, these matrices have full column-rank with probability $1$. We refer the reader for a detailed discussion on the rank of randomly generated matrices in a finite field to~\cite{bourgain2010singularity}.

\noindent {\bf Phase 2} [transmitting random linear combinations]: In this phase, transmitter ${\sf Tx}_i$ combines the bits in $Q_{i,1}$ and $Q_{i,2}$ to create $\tilde{Q}_i$ using the following equation.
\begin{align}
\tilde{Q}_i \overset{\triangle}= \mathbf{C}_{i,1} Q_{i,1} \oplus \mathbf{C}_{i,2} Q_{i,2}, \qquad i=1,2.
\end{align}

\begin{figure}[ht]
\centering
\includegraphics[height = 3.5 cm]{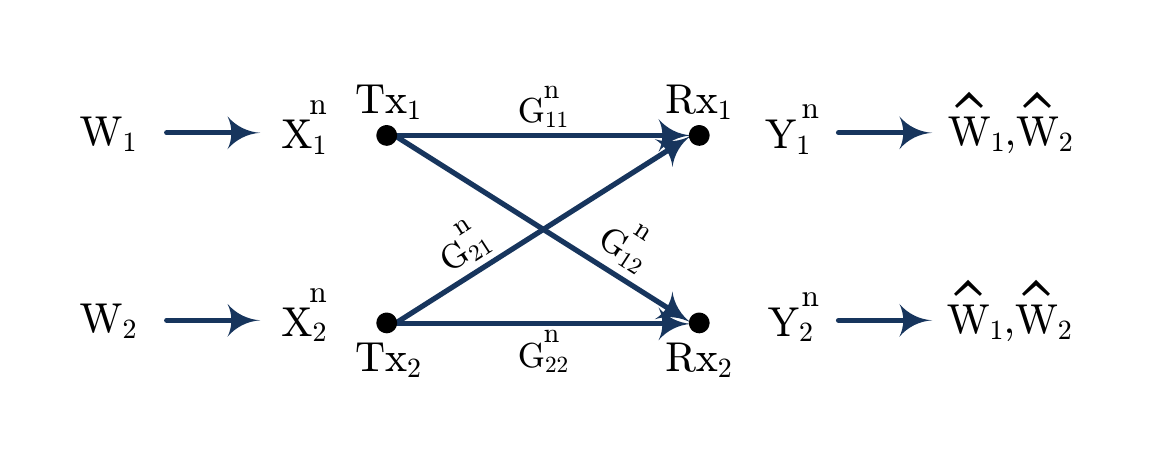}
\caption{Two-multicast network. Transmitter ${\sf Tx}_i$ wishes to reliably communicate message $\hbox{W}_i$ to both receivers, $i=1,2$. The capacity region with no or delayed CSIT is the same.\label{fig:two-multicast}}
\end{figure}

Then the goal is to provide the bits in $\tilde{Q}_1$ and $\tilde{Q}_2$ to \emph{both} receivers. The problem resembles a network with two transmitters and two receivers where each transmitter ${\sf Tx}_i$ wishes to communicate an independent message $\hbox{W}_i$ to {\it both} receivers as depicted in Fig.~\ref{fig:two-multicast}, $i=1,2$. The channel gain model is the same as described in Section~\ref{Section:Problem}. We refer to this problem as the  two-multicast problem. It is a straightforward exercise to show that for this problem, a rate-tuple of $\left( R_1, R_2 \right) = \left( \frac{3}{8}, \frac{3}{8} \right)$ is achievable. In other words, for fixed $\epsilon,\delta >0$, rate-tuple $\left( R_1, R_2 \right) = \left( \frac{3}{8}-\frac{\delta}{2} , \frac{3}{8}-\frac{\delta}{2} \right)$ is achievable with error less than or equal to $\epsilon$.

Fix $\epsilon,\delta >0$. Then, transmitters encode and communicate the bits in $\tilde{Q}_1$ and $\tilde{Q}_2$ using the achievability strategy of the two-multicast problem during Phase~2. This phase lasts for 
\begin{align}
\frac{\frac{2m}{3} + 8 m^{\frac{2}{3}}}{\frac{3}{4}-\delta}
\end{align} 
time instants. We assume $\tilde{Q}_1$ and $\tilde{Q}_2$ are decoded successfully at both receivers and no error has occurred.

\noindent {\bf Decoding}: At the end of Phase~2, receiver ${\sf Rx}_i$ removes the known bits from $Q_{\bar{i},1}$ and $Q_{\bar{i},2}$ (from (\ref{Eq:KnownBits}), we know that ${\sf Rx}_i$ has knowledge of at least $\frac{m}{3}$ bits). 

Thus after removing the known bits, receiver ${\sf Rx}_i$ has access to $\frac{m}{3} + 4 m^{\frac{2}{3}}$ random linear combinations of (at most) $\frac{m}{3} + 4 m^{\frac{2}{3}}$ unknown bits. Consequently, ${\sf Rx}_i$ can reconstruct all the bits in $Q_{\bar{i},1}$ and $Q_{\bar{i},2}$ with probability $1$ as $m \rightarrow \infty$. Then, receiver ${\sf Rx}_i$ uses the bits in $Q_{\bar{i},1}$ and $Q_{\bar{i},2}$ to remove the interference. Upon successfully removing interfering bits, the bits intended for ${\sf Rx}_i$ can be reconstructed from the available linear combinations. The reconstructing of the intended bits can be carried out error free with probability $1$ as $m \rightarrow \infty$. 

The total communication time is then equal to the length of Phase~1 plus the length of Phase~2. Thus when $\epsilon,\delta \rightarrow 0$, the total communication time is
\begin{align}
\frac{4}{3} m + m^{\frac{2}{3}} + \frac{4}{3} \left(  \frac{2m}{3} + 8 m^{\frac{2}{3}} \right) = \frac{20}{9} m + \frac{35}{3} m^{\frac{2}{3}}.
\end{align}

Hence, if we let $m \rightarrow \infty$, the decoding error probability at each phase of delivering the bits goes to zero exponentially, and we achieve a symmetric sum-rate of 
\begin{align}
R_1 = R_2 = \lim_{m \rightarrow \infty}{\frac{m}{\frac{20}{9} m + \frac{35}{3} m^{\frac{2}{3}}}} = 0.45 \raisebox{2pt}{.}
\end{align}

This completes the achievability proof for the corner point $\left( R_1, R_2 \right) = \left( 0.45, 0.45 \right)$.






\section{Discussion}
\label{Section:Discussion}

In this section, we discuss the problem of two-user erasure IC with local delayed CSIT given by View V.7, and then we try to understand the implications of our results in broader settings.

\subsection{Two-user Erasure IC with local delayed CSIT of View V.7}

Consider the two-user erasure IC with local delayed CSIT according to View V.7. We have
\begin{align} 
&\mathcal{S}_{{\sf Tx}_1} = \{ \left( 1, 1 \right), \left( 2, 1 \right), \left( 2, 2 \right) \} \quad \text{and} \nonumber \\
& \mathcal{S}_{{\sf Tx}_2} = \{ \left( 1, 1 \right), \left( 1, 2 \right), \left( 2, 2 \right) \}.
\end{align}
Thus writing the marginal distribution at receiver ${\sf Rx}_1$, we get
\begin{align}
\label{eq:MarginalView7Rx1}
& \Pr \left( Y_1^n, G^n | X_1^n, X_2^n \right) \nonumber \\
& = \left[ \frac{\Pr \left( G_{11}^n, G_{12}^n, G_{21}^n, G_{22}^n \right)}{\Pr \left( X_1^n,X_1^n \right)} \right] \nonumber \\
& \times \Pr \left( X_1^n | G_{11}^n, G_{21}^n, G_{22}^n \right) \Pr \left( X_2^n | G_{11}^n, G_{12}^n , G_{22}^n \right) \nonumber \\
& \mathbf{1}_{\left\{ Y_1^n = G_{11}^n X_1^n \oplus G_{21}^n X_2^n \right\}}.
\end{align}
Here, note that we can no longer use our trick in Section~\ref{Section:NoGain}. For instance, if we set
\begin{align}
\tilde{G}_{11}[t] = \tilde{G}_{12}[t],
\end{align}
then, we have changed the channel from ${\sf Tx}_2$'s point of view and thus, the marginal distributions cannot be preserved. 

On the other hand, as discussed in Section~\ref{Section:FullGain}, delayed knowledge of $G_{i\bar{i}}$ has an important role on the future decisions taken by ${\sf Tx}_i$, $i=1,2$. In fact, we cannot distinguish $Q_{i,1}$ from $Q_{i,2}$ without delayed knowledge of $G_{i\bar{i}}$, and thus, our achievability strategy cannot be utilized with local delayed CSIT of View V.7. 

In the absence of an achievability that goes beyond the capacity region with no CSIT, or a converse that matches that of no CSIT, the capacity region with local delayed CSIT of View V.7 remains open.

\subsection{$k$-user Erasure IC with delayed CSIT}

\begin{figure}[ht]
\centering
\includegraphics[height = 6cm]{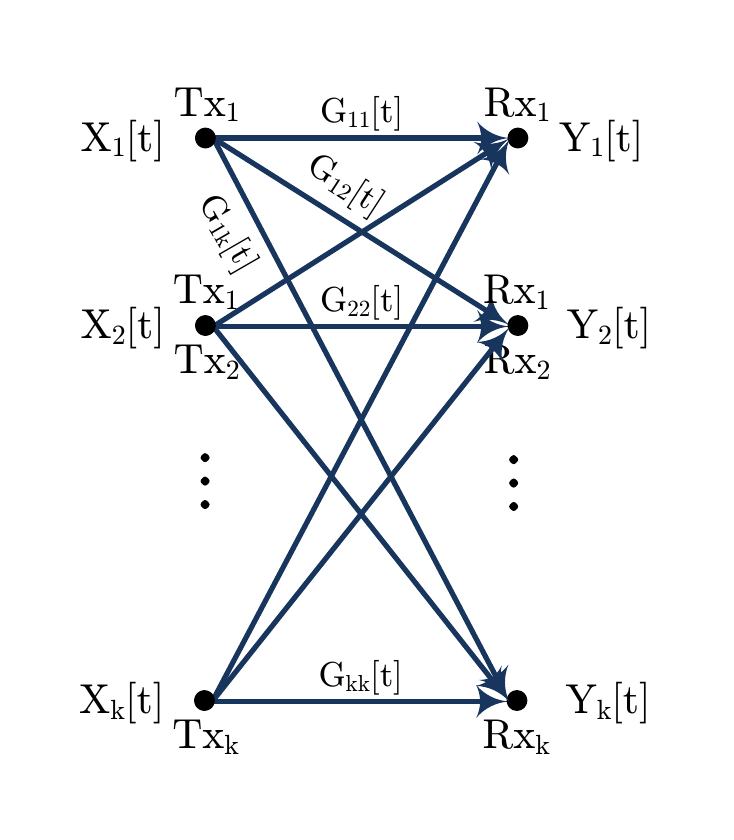}
\caption{\it $k$-user Erasure Interference Channel. The capacity region with global delayed CSIT is open.\label{Fig:KUser}}
\end{figure}

Here, we take the results and intuitions obtained for the two-user erasure IC and try to understand the implications in broader settings. We consider the capacity region of the $k$-user erasure IC (see Fig.~\ref{Fig:KUser}) and the degrees of freedom (DoF) region of the $k$-user Gaussian IC with Delayed CSIT. We denote the DoF region of the $k$-user Gaussian IC with global delayed CSIT by $\mathcal{D}_{k}$.

The intuition for the two-user erasure IC was that it is the responsibility of the transmitter who creates interference to resolve it. Characterizing $\mathcal{D}_{k}$ or the capacity region of the $k$-user erasure IC with global delayed CSIT are still open. However, there are several results that try to exploit the delayed knowledge of the channel state information for the achievability purposes in the context of $k$-user Gaussian IC (\emph{e.g.}, see~\cite{Jafar_Retrospective,abdoli2011degrees} and references therein). In~\cite{ghasemi2011interference}, authors have shown that such gains can be also obtained if each transmitter is only aware of the channel gains of the outgoing links from itself with delay. This result matches our intuition for the two-user erasure IC. However, in the lack of a tight outer-bound, a firm conclusion cannot be made.


\section{Conclusion and Future Directions}
\label{Section:Conclusion}

\begin{figure*}[t]
\centering
\includegraphics[height = 12cm]{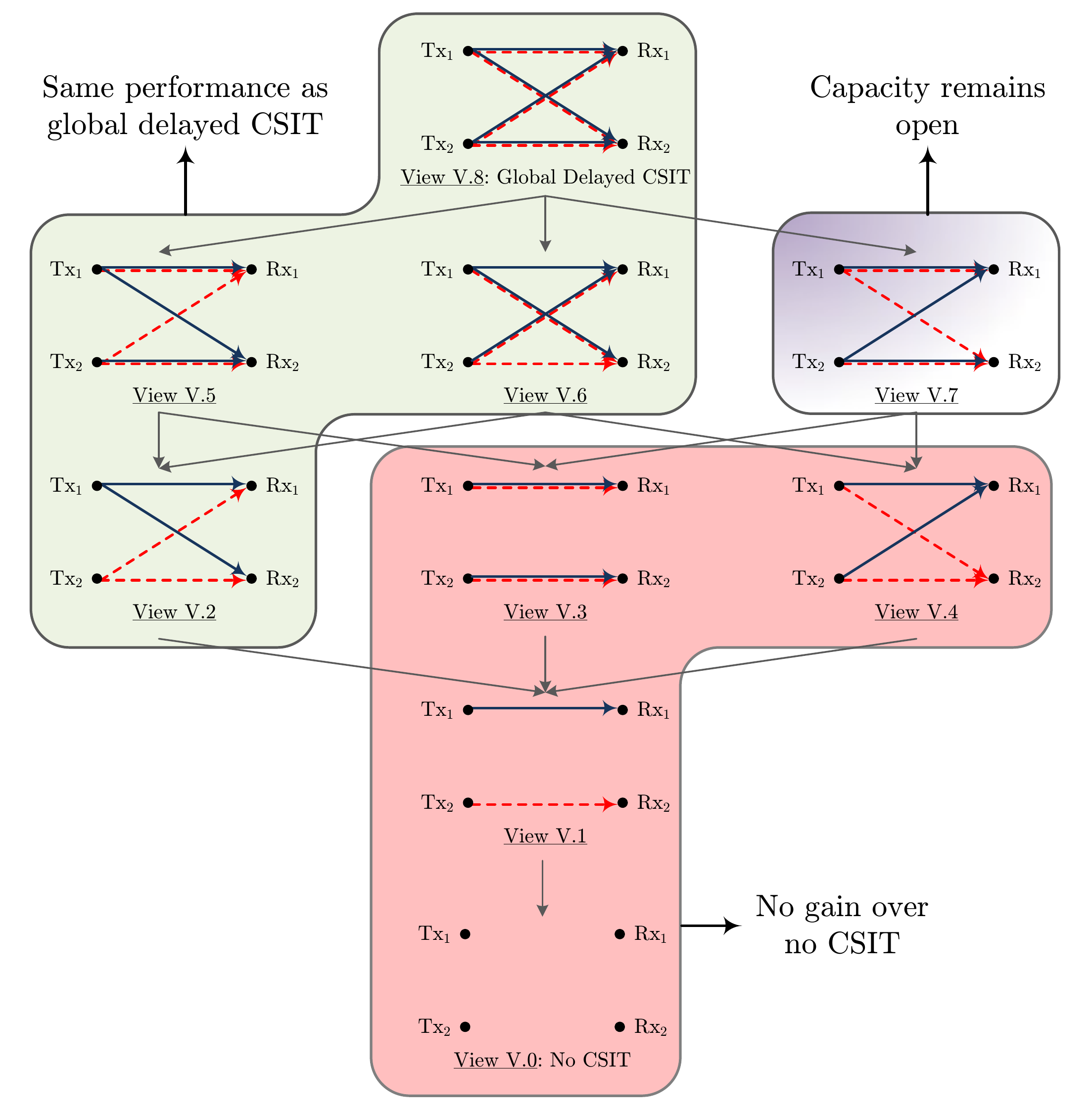}
\caption{\it Summary of the main results. Capacity region with local delayed CSIT of Views V.1, V.3, and V.4 coincides with no CSIT case (V.0); while capacity region with local delayed CSIT of Views V.2, V.5, and V.6 matches that of global delayed CSIT (V.8). The capacity region with local delayed CSIT of View V.7 remains open.\label{Fig:Summary}}
\end{figure*}

We studied the capacity region of the two-user Binary Fading Interference Channel with \emph{local} delayed channel state information at the transmitters. We showed that in order to achieve the performance of \emph{global} delayed CSIT, it suffices that each transmitter has only access to
the delayed knowledge of its {\it outgoing} links. We also identified the cases in which local delayed CSIT does not provide any gain over the no knowledge assumption. Fig.~\ref{Fig:Summary}, summarizes our main results.

As discussed in Section~\ref{Section:Discussion}, an interesting future direction is to extend the result to the $k$-user Binary Fading Interference Channel and see whether the delayed knowledge of the outgoing links suffices to achieve the capacity with global delayed CSIT. This result, if true, would shed light on finally solving the capacity region (or DoF region) of $k$-user interference channels with delayed CSIT. Another direction, would be to extend the current results to two-user Rayleigh fading interference channels (as opposed to the binary fading model). Recently, a new direction was introduced in~\cite{vahid2016does} where spatial correlation between channels were considered. Implications of local delayed CSIT in that setting has great practical importance. 

\appendices

\section{More Discussion on Theorem~\ref{THM:NoGain}}
\label{Appendix:NewAppendix}

Define
\begin{align}
\mathcal{E}_{\hbox{W}_1} \overset{\triangle}= \left\{ \left( W_2, G^n \right) \text{~s.t.~} \widehat{\hbox{W}}_1 \neq \hbox{W}_1 \right\},
\end{align} 
and
\begin{align}
\mathcal{E}^1_{\hbox{W}_1} \overset{\triangle}= \left\{ \left( X_1^n, G_{11}^n, G_{21}^n \right) \text{~s.t.~} \widehat{\hbox{W}}_1 \neq \hbox{W}_1 \right\}, \nonumber \\
\mathcal{E}^2_{\hbox{W}_1} \overset{\triangle}= \left\{ \left( X_2^n, G_{12}^n, G_{22}^n \right) \text{~s.t.~} \widehat{\hbox{W}}_1 \neq \hbox{W}_1 \right\}.
\end{align} 
Then, we have
\begin{align}
\lambda_{1,n} &= \sum_{w_1}{\Pr\left( \hbox{W}_1 = w_1 \right) \Pr\left( \mathcal{E}_{\hbox{W}_1 = w_1} \right)} \nonumber \\
&\overset{(\ref{eq:MarginalView4Rx1})}= \sum_{w_1}{\Pr\left( \hbox{W}_1 = w_1 \right) \left[ \Pr\left( \mathcal{E}^1_{\hbox{W}_1 = w_1} \right) + \Pr\left( \mathcal{E}^2_{\hbox{W}_1 = w_1} \right) \right]} \nonumber \\
&= \sum_{w_1}{\Pr\left( \hbox{W}_1 = w_1 \right) \left[ \Pr\left( \tilde{\mathcal{E}}^1_{\hbox{W}_1 = w_1} \right) + \Pr\left( \tilde{\mathcal{E}}^2_{\hbox{W}_1 = w_1} \right) \right]} \nonumber \\
& \geq \sum_{w_1}{\Pr\left( \hbox{W}_1 = w_1 \right) \Pr\left( \tilde{\mathcal{E}}_{\hbox{W}_1 = w_1} \right)} \nonumber \\
& = \tilde{\lambda}_{1,n}, 
\end{align}
where $\tilde{\mathcal{E}}_{\hbox{W}_1}$, $\tilde{\mathcal{E}}^j_{\hbox{W}_1}$, and $\tilde{\lambda}_{1,n}$ are respectively the counterparts of $\mathcal{E}_{\hbox{W}_1}$, $\mathcal{E}^j_{\hbox{W}_1}$, and $\lambda_{1,n}$ for the new channel, $j=1,2$.

Similarly, we can start with $\tilde{\lambda}_{1,n}$ and use the fact that
\begin{align}
\Pr\left( \tilde{\mathcal{E}}_{\hbox{W}_1 = w_1} \right) & \leq \Pr\left( \tilde{\mathcal{E}}^1_{\hbox{W}_1 = w_1} \right) + \Pr\left( \tilde{\mathcal{E}}^2_{\hbox{W}_1 = w_1} \right) \nonumber \\
& \leq 2\Pr\left( \tilde{\mathcal{E}}_{\hbox{W}_1 = w_1} \right),
\end{align} 
to show that
\begin{align}
\lambda_{1,n} \rightarrow 0 \Leftrightarrow \tilde{\lambda}_{1,n} \rightarrow 0.
\end{align} 
Therefore, we conclude that
\begin{align}
\mathcal{C}\left( \text{V.}4 \right) \equiv \tilde{\mathcal{C}}\left( \text{V.}4 \right).
\end{align} 


\section{Transmission Strategy for the Corner Point $\left( 0.375, 0.5 \right)$}
\label{Appendix:Corner}

We now provide the achievability strategy for the corner point
\begin{align}
\left( R_1, R_2 \right) = \left( \frac{3}{8}, \frac{1}{2} \right) \raisebox{2pt}{.}
\end{align}

To achieve this corner point, new challenges arise which are due to the asymmetry of the rates. In this case, ${\sf Tx}_2$ (the primary user) communicates at the full rate of $0.5$ while ${\sf Tx}_1$ (the secondary user) communicates at a lower rate and tries to coexist with the primary user. In fact ${\sf Tx}_1$ has to take more responsibility in dealing with interference at {\it both} receivers.  The proposed transmission strategy consists of four phases as described below. We assume that ${\sf Tx}_1$ has $\frac{3}{4}m$ bits to communicate while ${\sf Tx}_2$ has $m$ bits. We show that all bits can be delivered in $2m + O\left( m^{2/3} \right)$ time instants with vanishing error probability as $m \rightarrow \infty$. This immediately implies the achievability of the corner point $\left( 0.375, 0.5 \right)$.

\noindent {\bf Phase 1}: This phase is similar to Phase 1 of the achievability of the optimal sum-rate point $\left( 0.45, 0.45 \right)$. The main difference is due to the fact that the transmitters have unequal number of bits at the start. In Phase $1$, ${\sf Tx}_1$ (the secondary user) transmits all its initial bits while ${\sf Tx}_2$ (the primary user) only transmits a fraction of its initial bits. Transmitter two postpones the transmission of its remaining bits to Phase 2.

At the beginning of the communication block, we assume that each transmitter has $\frac{3}{4}m$ bits in queue $Q_{i \rightarrow i}$ (the initial state of the bits), $i=1,2$. At each time instant $t$, ${\sf Tx}_i$ sends out a bit from $Q_{i \rightarrow i}$, and this bit will either stay in the initial queue or a transition to one of the following possible queues will take place according to the description in Table~\ref{Table:View2}. If at time instant $t$, $Q_{i \rightarrow i}$ is empty, then ${\sf Tx}_i$, $i=1,2$, remains silent until the end of Phase 1.

Phase~$1$ continues for 
\begin{align}
m + m^{\frac{2}{3}}
\end{align}
time instants, and if at the end of this phase, either of the queues $Q_{i \rightarrow i}$ is not empty, we declare error type-I and halt the transmission.

Assuming that the transmission is not halted, let $N_{i,1}$ and $N_{i, 2}$, $i=1,2$, denote the number of bits in queues $Q_{i,1}$ and $Q_{i,2}$ respectively at the end of Phase~$1$. The transmission strategy will be halted and an error type-II will occur, if any of the following events happens.
\begin{align}
& N_{i,1} > \mathbb{E}[N_{i,1}] + 2 m^{\frac{2}{3}} \overset{\triangle}= n_{i,1}, \quad i=1,2; \nonumber \\
& N_{i,2} > \mathbb{E}[N_{i,2}] + 2 m^{\frac{2}{3}} \overset{\triangle}= n_{i,2}, \quad i=1,2.
\end{align}

From basic probability, we have
\begin{align}
\mathbb{E}[N_{i,1}] = \mathbb{E}[N_{i,2}] = \frac{m}{4},
\end{align}
so that
\begin{align}
n_{i,1} =  n_{i,2} = \frac{m}{4} + 2 m^{\frac{2}{3}}.
\end{align}
At the end of Phase $1$, for $i=1,2$, we add $0$'s (if necessary) in order to make queues $Q_{i,1}$ and $Q_{i,2}$ of size equal to $n_{i,1}$ and $n_{i,2}$ respectively.

Since channel gains are distributed independently, statistically half of the bits in $Q_{i,1}$ and half of the bits in $Q_{i,2}$ are known to ${\sf Rx}_{\bar{i}}$, $i=1,2$. Denote the number of bits in $Q_{i,j}$ known to ${\sf Rx}_{\bar{i}}$ by
\begin{align}
N_{i,j|{\sf Rx}_{\bar{i}}}, \qquad i,j \in \{ 1, 2 \}.
\end{align}
At the end of communication, if we have
\begin{align}
N_{i,j|{\sf Rx}_{\bar{i}}} < \frac{1}{2} n_{i,j} - m^{\frac{2}{3}}, \qquad i,j \in \{ 1, 2 \},
\end{align}
we declare error type-III.

Moreover, we note that statistically for every two bits in $Q_{\bar{i},1}$, a bit in $Q_{i,1}$ was transmitted simultaneously with one of them. Denote the number of bits in $Q_{i,1}$ that were transmitted simultaneously with a bit in $Q_{\bar{i},1}$ by
\begin{align}
N_{i \rightarrow \bar{i},1}, \qquad i = 1, 2.
\end{align}
At the end of communication, if we have
\begin{align}
N_{i \rightarrow \bar{i},1} < \frac{1}{2} n_{i,1} - m^{\frac{2}{3}}, \qquad i = 1, 2,
\end{align}
we declare error type-IV. Note that transmitters cannot detect error type-III or error type-IV, but receivers have sufficient information to do so. 

Using the Bernstein inequality, we can show that the probability of errors of types I, II, III, and IV decreases exponentially with $m$. For the rest of this subsection, we assume that Phase~1 is completed and no error has occurred.


\noindent {\bf Phase 2} [transmission of new bits vs interference management]: In this phase, the primary user ${\sf Tx}_2$ transmits its remaining initial bits while the secondary user ${\sf Tx}_1$ tries to resolve as much interference as it can and deliver some of its bits in $Q_{1,1}$. To do so, the secondary user sends some of its bits in $Q_{1,1}$ at a rate low enough such that both receivers can decode and remove them regardless of what the primary transmitter does. Note that $1/4$ of the time, each receiver obtains an interference-free signal from the secondary transmitter, hence, the secondary transmitter can take advantage of these time instants to deliver its bits during Phase~2. 

\begin{table*}[t]
\caption{Transition of the bits from $Q_{2 \rightarrow 2}$ during Phase~2.}
\centering
\begin{tabular}{| c | c | c | c | c | c |}
\hline
case ID		 & channel realization    & state transition  & case ID		 & channel realization    & state transition \\
					 & at time instant $n$    &                   & 					 & at time instant $n$    &                  \\ [0.5ex]
\hline

\raisebox{30pt}{$1$}    &    \includegraphics[height = 2.5cm]{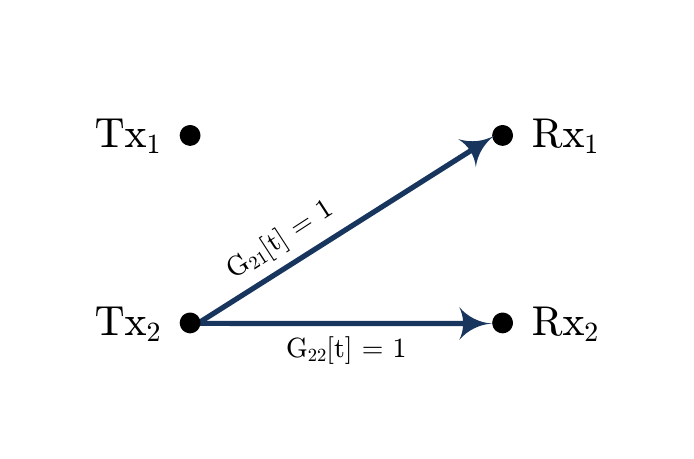}     &  \raisebox{30pt}{$a \rightarrow Q_{1,1}$}  &  \raisebox{18pt}{$3$}    &    \includegraphics[height = 2.5cm]{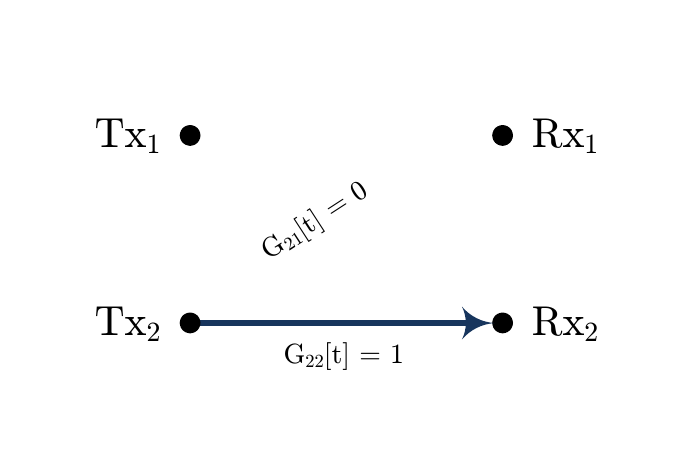}     &   \raisebox{18pt}{$a \rightarrow Q_{1 \rightarrow F}$} \\

\hline

\raisebox{30pt}{$2$}    &    \includegraphics[height = 2.5cm]{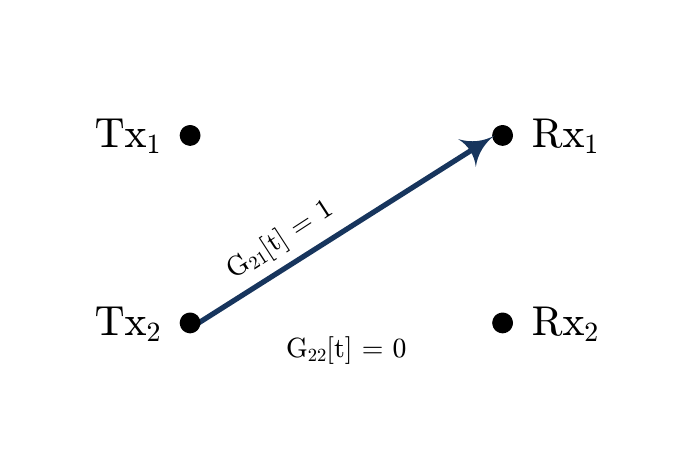}     &  \raisebox{30pt}{$a \rightarrow Q_{1,2}$} &  \raisebox{18pt}{$4$}    &    \includegraphics[height = 2.5cm]{FiguresPDF/ICD.pdf}     &   \raisebox{18pt}{$a \rightarrow Q_{1 \rightarrow 1 }$} \\

\hline

\end{tabular}
\label{Table:View2Corner}
\end{table*}

Transmitter ${\sf Tx}_1$ creates a matrix $\mathbf{C}_{1,1}$ of size $\left( \frac{1}{12}m+m^{2/3} \right) \times \left( \frac{m}{4} + 2 m^{\frac{2}{3}} \right)$, where the entries of this matrix are drawn from i.i.d. $\mathcal{B}(0.5)$ distribution. We assume that this matrix is generated prior to communication and is shared with receivers. Then ${\sf Tx}_1$ creates $\left( \frac{1}{12}m+m^{2/3} \right)$ bits by multiplying matrix $\mathbf{C}_{1,1}$ and the bits in $Q_{1,1}$. Using point-to-point erasure code of rate $1/4$, transmitter ${\sf Tx}_1$ encodes bits $\mathbf{C}_{1,1} Q_{1,1}$ and communicates them during Phase~2. We note that due to the chosen rate (\emph{i.e.} $1/4$) as $m \rightarrow \infty$, each receiver can decode bits $\mathbf{C}_{1,1} Q_{1,1}$ with vanishing error probability.

Transmitter ${\sf Tx}_2$ places its remaining $\frac{1}{4}m$ bits in queue $Q_{2 \rightarrow 2}$ (the initial state of the bits). At each time instant $t$ of Phase~2, ${\sf Tx}_2$ sends out a bit from $Q_{2 \rightarrow 2}$, and this bit will either stay in the initial queue or a transition to a new queue will take place according to the description in Table~\ref{Table:View2Corner}. Note that here, since the signal of ${\sf Tx}_1$ can be decoded first, we simply consider the bits of ${\sf Tx}_2$ that were transmitted in case $1$ (see Table~\ref{Table:View2Corner}) to be delivered. At the end of Phase~2, we update the value of $n_{2,2}$ as
\begin{align}
n_{2,2} = \frac{m}{3} + 3 m^{\frac{2}{3}}.
\end{align}



\noindent {\bf Phase 3} [encoding and mixing interfering bits]: Transmitter ${\sf Tx}_1$ creates two matrices: $\mathbf{C}_{1,2}$ of size $\left( \frac{m}{6} + 2 m^{\frac{2}{3}} \right) \times \left( \frac{m}{4} + 2 m^{\frac{2}{3}} \right)$ and $\mathbf{C}_{1,3}$ of size $\left( \frac{m}{4} + 2 m^{\frac{2}{3}} \right) \times \left( \frac{m}{4} + 2 m^{\frac{2}{3}} \right)$, where entries to each matrix are drawn from i.i.d. $\mathcal{B}(0.5)$ distribution. We assume that the matrices are generated prior to communication and are shared with receivers. 

Transmitter ${\sf Tx}_1$ creates
\begin{align} 
\tilde{Q}_{1,1} = \mathbf{C}_{1,2} Q_{1,1}, \nonumber\\
\tilde{Q}_{1,2} = \mathbf{C}_{1,3} Q_{1,2}. 
\end{align}
Then ${\sf Tx}_1$ encodes bits in $\tilde{Q}_{1,1}$ using a point-to-point erasure code of rate $1/4$ denoted by $\hat{Q}_{1,1}$ and encodes bits in $\tilde{Q}_{1,2}$ using a point-to-point erasure code of rate $1/2$ denoted by $\hat{Q}_{1,2}$. Transmitter ${\sf Tx}_1$ communicates\footnote{$,^3$ The two sequences are not of equal length, we can simply add deterministic number of zeros to $\hat{Q}_{1,2}$ to make the two sequences of equal length} $\hat{Q}_{1,1} \oplus \hat{Q}_{1,2}$ during Phase~3.

At the same time, transmitter ${\sf Tx}_2$ creates two matrices: $\mathbf{C}_{2,1}$ of size $\left( \frac{m}{8} + 2 m^{\frac{2}{3}} \right) \times \left( \frac{m}{4} + 2 m^{\frac{2}{3}} \right)$ and $\mathbf{C}_{2,2}$ of size $\left( \frac{m}{3} + 2 m^{\frac{2}{3}} \right) \times \left( \frac{m}{3} + 2 m^{\frac{2}{3}} \right)$, where entries to each matrix are drawn from i.i.d. $\mathcal{B}(0.5)$ distribution. We assume that the matrices are generated prior to communication and are shared with receivers.

Transmitter ${\sf Tx}_2$ creates
\begin{align} 
\tilde{Q}_{2,1} = \mathbf{C}_{2,1} Q_{2,1}, \nonumber\\
\tilde{Q}_{2,2} = \mathbf{C}_{2,2} Q_{2,2}. 
\end{align}
Then ${\sf Tx}_2$ encodes bits in $\tilde{Q}_{2,1}$ using a point-to-point erasure code of rate $1/4$ denoted by $\hat{Q}_{2,1}$ and encodes bits in $\tilde{Q}_{2,2}$ using a point-to-point erasure code of rate $1/2$ denoted by $\hat{Q}_{2,2}$. Transmitter ${\sf Tx}_2$ communicates$^3$ $\hat{Q}_{2,1} \oplus \hat{Q}_{2,2}$ during Phase~3.

\noindent {\bf Decoding}: Upon completion of the third phase, we show that each receiver has gathered enough linear equations to decode all bits in $Q_{1,1}, Q_{1,2}$ and $Q_{2,2}$. Receiver ${\sf Rx}_i$ first removes the known bits from $Q_{\bar{i},1}$ and $Q_{\bar{i},2}$, $i=1,2$.

Then, each receiver has $$\underbrace{\left( \frac{m}{12} + 2 m^{\frac{2}{3}} \right)}_{\mathbf{C}_{1,1} Q_{1,1}} + \underbrace{\left( \frac{m}{6} + 2 m^{\frac{2}{3}} \right)}_{\mathbf{C}_{1,2} Q_{1,1}} = \frac{m}{4} + 2 m^{\frac{2}{3}}$$ randomly generated equations of $\frac{m}{4} + 2 m^{\frac{2}{3}}$ bits in $Q_{1,1}$. Thus, both receivers can recover the bits in $Q_{1,1}$ with vanishing error probability as $m \rightarrow \infty$. Similarly, receivers have sufficient information to recover bits in $Q_{1,2}$ and $Q_{2,2}$ 

As opposed to other states, not all bits in $Q_{2,1}$ are provided to the receivers by ${\sf Tx}_2$. However, transmitter ${\sf Tx}_2$ is not required to provide all bits in $Q_{2,1}$ to both receivers. The reason is that, once $Q_{1,1}$ is known at the receivers, statistically half of the bits in $Q_{2,1}$ can be reconstructed at each receiver. Therefore, transmitter ${\sf Tx}_2$ needs to provide half of the bits in $Q_{2,1}$ to the receivers. That is why in Phase~3, we chose $\mathbf{C}_{2,1}$ to have size $\left( \frac{m}{8} + 2 m^{\frac{2}{3}} \right) \times \left( \frac{m}{4} + 2 m^{\frac{2}{3}} \right)$ in lieu of $\left( \frac{m}{4} + 2 m^{\frac{2}{3}} \right) \times \left( \frac{m}{4} + 2 m^{\frac{2}{3}} \right)$.

We therefore conclude that each receiver can recover its intended bits with vanishing error probability as $m \rightarrow \infty$ in a total of
\begin{align}
2m+O\left( m^{2/3} \right)
\end{align}
time instants.

Hence, if we let $m \rightarrow \infty$, the decoding error probability goes to zero exponentially, and we achieve rate-tuple 
\begin{align}
\left( R_1, R_2 \right) = \left( \frac{3}{8}, \frac{1}{2} \right) \raisebox{2pt}{.}
\end{align}

Similarly, we can achieve the corner point
\begin{align}
\left( R_1, R_2 \right) = \left( \frac{1}{2}, \frac{3}{8} \right) \raisebox{2pt}{.}
\end{align}
Together with the results of Section~\ref{Section:FullGain}, we conclude that $\mathcal{C}\left( \text{V.}8 \right)$ is achievable with local delayed CSIT of View V.2.

\section{Transmission Strategy for $0 \leq p \leq 1$}
\label{Appendix:GeneralP}

The reason we considered $p =0.5$ in Section~\ref{Section:FullGain} was to remain focused on the impact of local delayed CSIT on the capacity region rather than getting involved in the details of the transmission strategy.

Here, we describe the changes needed in the transmission strategy when considering $0 \leq p \leq 1$. The core structure of the achievability strategy remains the same as what we discussed in Section~\ref{Section:FullGain}. However, modifications are needed to ensure optimal performance. 

Consider the maximum symmetric sum-rate point as given by
\begin{align}
\label{Eq:MaxSumRate}
R_1 = R_2 = \min\left\{ p, \frac{\beta\left( 1 - q^2 \right)}{1+\beta} \right\},
\end{align}
where
\begin{align}
\beta = 2-p.
\end{align}
The achievability strategy for the other corner points, \emph{i.e.}
\begin{align}
\left( R_i, R_{\bar{i}} \right) = \left( \min\left\{ p, pq(1+q)\right\}, p \right), \qquad i=1,2,
\end{align}
follows similar modifications when compared to the strategy given for $p=0.5$ in Appendix~\ref{Appendix:Corner}. We note that the capacity region is the convex hull of the aforementioned corner points.

Suppose each transmitter wishes to communicate $m$ bits to its intended receiver. We need to show that this task can be accomplished (with vanishing error probability as $m \rightarrow \infty$) in
\begin{align}
\max\left\{ \frac{1}{p}, \frac{1+\beta}{\beta\left( 1 - q^2 \right)} \right\} m + \mathcal{O}\left( m^{\frac{2}{3}} \right)
\end{align}
time instants. 

In Section~\ref{Section:FullGain}, we considered $p =0.5$ and that implies
\begin{align}
\mathbb{E}[N_{i,1}] = \mathbb{E}[N_{i,2}].
\end{align}

However, when $p \neq 0.5$ the above inequality no longer holds. Below, we describe other coding opportunities that were not needed in Section~\ref{Section:FullGain}.
\begin{enumerate}

\item Suppose $p > 0.5$, then we have $\mathbb{E}[N_{i,1}] > \mathbb{E}[N_{i,2}]$. After combining the bits in $Q_{i,1}$ and $Q_{i,2}$, one naive solution would be to treat the remaining bits in $Q_{i,1}$ as bits of common interest. However, we can improve upon that scheme as described below.

Suppose each one of the transmitters sends three data bits as depicted in Fig.~\ref{fig:opp-commoncodingswapped}. We observe that providing $a_9 \oplus a_{10}$ and $b_9 \oplus b_{11}$ to both receivers is sufficient to decode the bits. For instance, if ${\sf Rx}_1$ is provided with $a_9 \oplus a_{10}$ and $b_9 \oplus b_{11}$, then it will use $a_{10}$ to decode $a_9$, from which it can obtain $b_9$; then using $b_9$ and $b_9 \oplus b_{11}$, it gains access to $b_{11}$; finally using $b_{11}$, it can decode $a_{11}$ from $a_{11} \oplus b_{11}$. Thus, linear combination $a_9 \oplus a_{10}$ available at ${\sf Tx}_1$, and linear combination $b_9 \oplus b_{11}$ available at ${\sf Tx}_2$, are bits of common interest and can be transmitted to both receivers simultaneously in the efficient two-multicast problem. We note that the bits of ${\sf Tx}_1$ in Fig.~\ref{fig:opp-commoncodingswapped}(a) and Fig.~\ref{fig:opp-commoncodingswapped}(b) fall in $Q_{1,1}$ and the bits of ${\sf Tx}_2$ in Fig.~\ref{fig:opp-commoncodingswapped}(a) and Fig.~\ref{fig:opp-commoncodingswapped}(c) fall in $Q_{2,1}$. Thus, we can further combine the bits in $Q_{i,1}$ to improve the achievable rate region.

\begin{figure}[h]
\centering
\subfigure[]{\includegraphics[height = 2.5 cm]{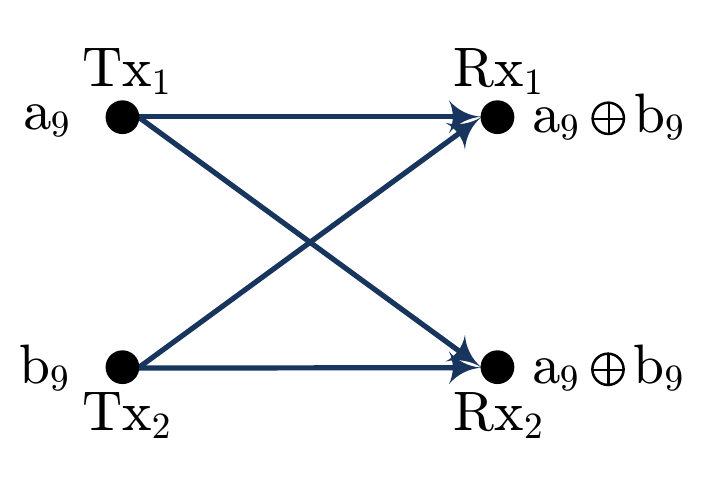}}
\hspace{0.1 in}
\subfigure[]{\includegraphics[height = 2.5 cm]{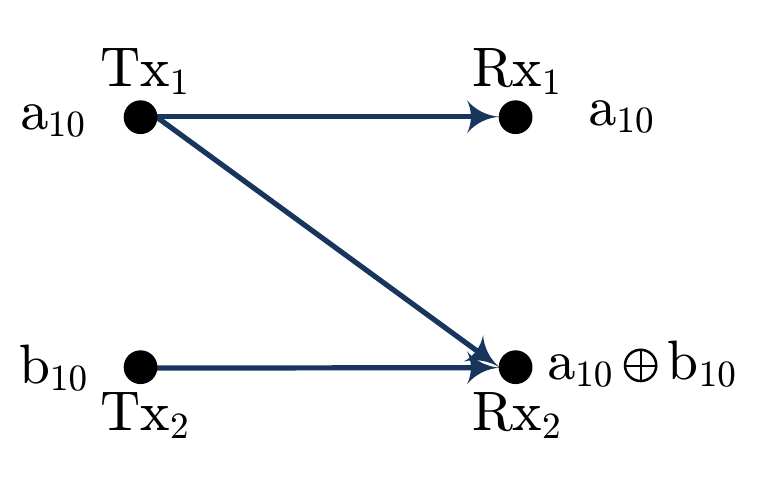}}
\subfigure[]{\includegraphics[height = 2.5 cm]{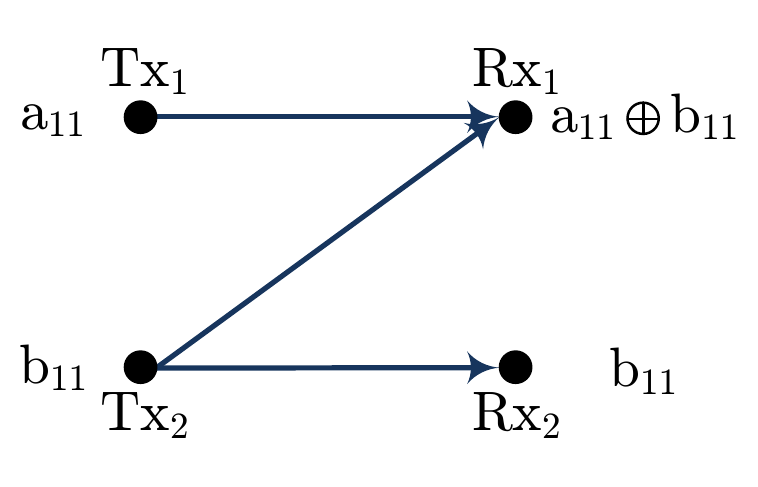}}
\caption{\it Providing $a_9 \oplus a_{10}$ and $b_9 \oplus b_{11}$ to both receivers is sufficient to decode the bits.\label{fig:opp-commoncodingswapped}}
\end{figure}

\item Suppose $p > 0.5$, then we have $\mathbb{E}[N_{i,1}] > \mathbb{E}[N_{i,2}]$. After combining the bits in $Q_{i,1}$ and $Q_{i,2}$, sufficient number of linear combinations of the remaining bits in $Q_{i,2}$ should be created and communicated using a point-to-point erasure code of rate $p$.

\end{enumerate}

It is important to realize that the coding opportunities described above can be identified using the local delayed CSIT. What remains is to pick the right number of linear combinations to communicate in each step of the transmission strategy. It is a straightforward exercise to use the results of~\cite{AlirezaBFICDelayed} to deduce the right number of linear combinations needed in each step of our strategy with local delayed CSIT.

\bibliographystyle{ieeetr}
\bibliography{bib_FBBudget}

\begin{IEEEbiography} [{\includegraphics[width=25mm]{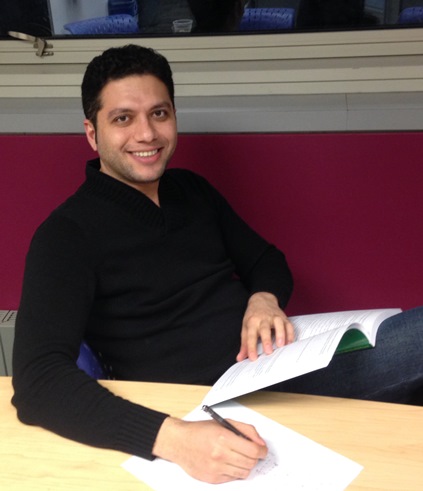}}] {Alireza Vahid} received the B.Sc. degree in electrical engineering from Sharif University of Technology, Tehran, Iran, in 2009, and the M.Sc. degree and Ph.D. degree in electrical and computer engineering both from Cornell University, Ithaca, NY, in 2012 and 2015 respectively. As of September 2014, he is a postdoctoral scholar at Information Initiative at Duke University, Durham, NC. His research interests include network information theory, wireless communications, statistics and machine learning.

Dr. Vahid received the 2015 Outstanding PhD Thesis Research Award at Cornell University. He also received the Director's Ph.D. Teaching Assistant Award in 2010, Jacobs Scholar Fellowship in 2009, and Qualcomm Innovation Fellowship in 2013.

\end{IEEEbiography}

\begin{IEEEbiography} [{\includegraphics[height=32mm]{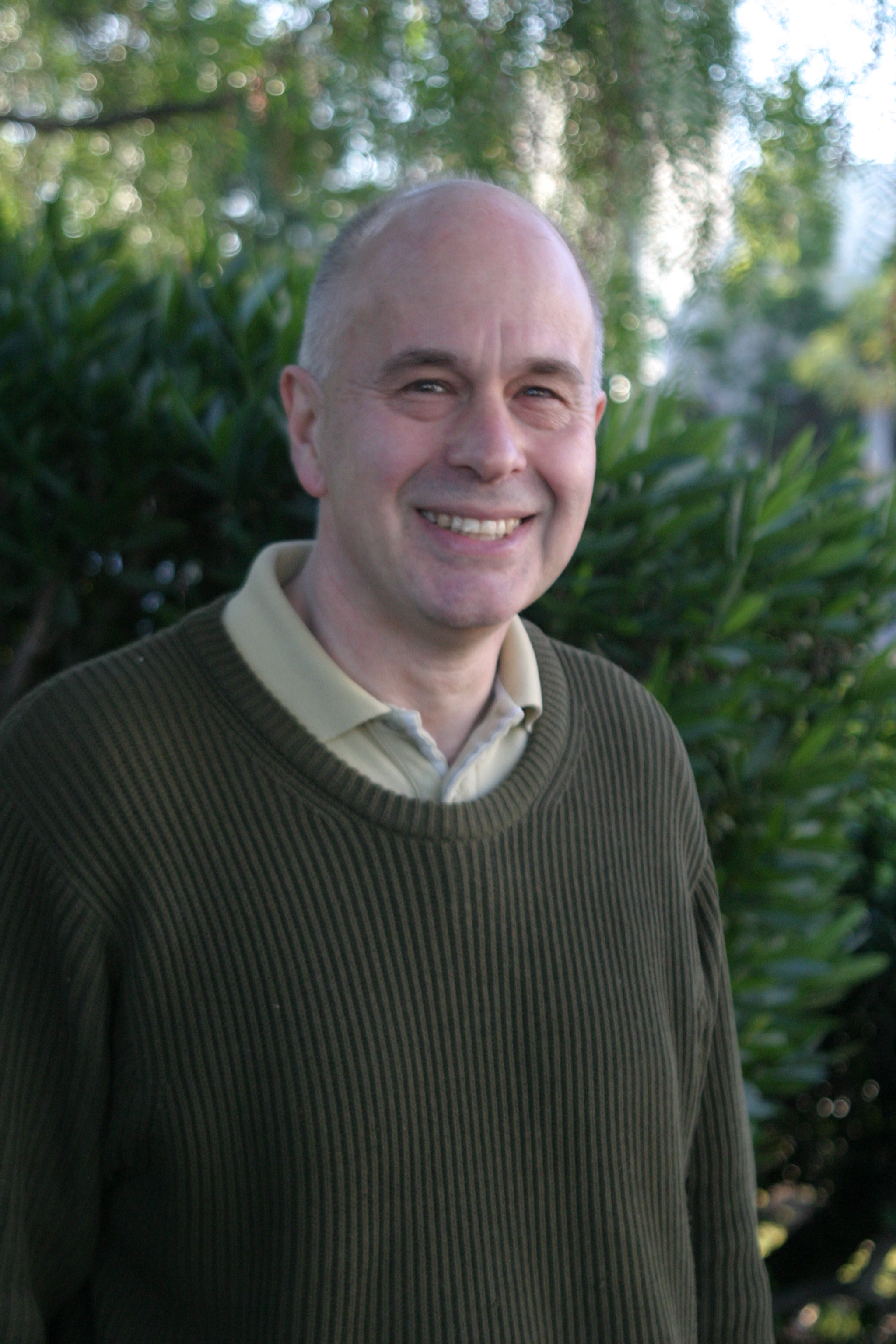}}] {Robert Calderbank} (M’89 – SM’97 – F’98) received the BSc degree in 1975 from Warwick University, England, the MSc degree in 1976 from Oxford University, England, and the PhD degree in 1980 from the California Institute of Technology, all in mathematics.

Dr. Calderbank is Professor of Electrical Engineering at Duke University where he now directs the Information Initiative at Duke (\emph{i}iD) after serving as Dean of Natural Sciences (2010-2013). Dr. Calderbank was previously Professor of Electrical Engineering and Mathematics at Princeton University where he directed the Program in Applied and Computational Mathematics. Prior to joining Princeton in 2004, he was Vice President for Research at AT\&T, responsible for directing the first industrial research lab in the world where the primary focus is data at scale. At the start of his career at Bell Labs, innovations by Dr. Calderbank were incorporated in a progression of voiceband modem standards that moved communications practice close to the Shannon limit. Together with Peter Shor and colleagues at AT\&T Labs he showed that good quantum error correcting codes exist and developed the group theoretic framework for quantum error correction. He is a co-inventor of space-time codes for wireless communication, where correlation of signals across different transmit antennas is the key to reliable transmission.

Dr. Calderbank served as Editor in Chief of the IEEE TRANSACTIONS ON INFORMATION THEORY from 1995 to 1998, and as Associate Editor for Coding Techniques from 1986 to 1989. He was a member of the Board of Governors of the IEEE Information Theory Society from 1991 to 1996 and from 2006 to 2008. Dr. Calderbank was honored by the IEEE Information Theory Prize Paper Award in 1995 for his work on the $Z_4$ linearity of Kerdock and Preparata Codes (joint with A.R. Hammons Jr., P.V. Kumar, N.J.A. Sloane, and P. Sole), and again in 1999 for the invention of space-time codes (joint with V. Tarokh and N. Seshadri). He has received the 2006 IEEE Donald G. Fink Prize Paper Award, the IEEE Millennium Medal, the 2013 IEEE Richard W. Hamming Medal, the 2015 Shannon Award, and he was elected to the US National Academy of Engineering in 2005.

\end{IEEEbiography}

\end{document}